# Characterization of hexagonal boron nitride layers on nickel surfaces by low-energy electron microscopy

P. C. Mende,[1] Q. Gao,[1] A. Ismach,[2] H. Chou,[2] M. Widom,[1] R. Ruoff,[2] L. Colombo,[3] and R. M. Feenstra[1,*]

[1]Dept. Physics, Carnegie Mellon University, Pittsburgh, PA 15213
[2]Dept. Mechanical Engineering, University of Texas at Austin, Austin, TX 78712
[3]Texas Instruments, Advanced Technology Development, Dallas, TX 75243

## Abstract

The thickness and interfacial geometry of hexagonal boron nitride (hBN) films grown by chemical vapor deposition on polycrystalline nickel foils is studied using low-energy electron microscopy (LEEM). The reflectivity of the electrons, measured over an energy range of 0 – 20 eV, reveals distinct minima and maxima. The measured data is compared with simulations based on a first-principles description of the electronic structure of the material. From this comparison, the number of hBN layers and the separation between the lowest hBN layer and the nickel surface is deduced. The coupling of interlayer states of the hBN to both image-potential and Shockley-type surface states of the nickel is discussed, and the dependence of the reflectivity spectra on the surface orientation of nickel grains is examined.

## I. Introduction

For many reasons, hexagonal boron nitride (hBN) is an ideal candidate for graphene-based vertical and horizontal heterostructures: hBN and graphene have a small lattice mismatch (< 2%); it is atomically flat, unlike $SiO_2$; it is chemically inert; and it produces no charge traps, again, unlike $SiO_2$.[1,2] Devices based on vertical heterostructures have been predicted to have unique properties[3], properties which have been realized recently in devices incorporating exfoliated hBN[4]. Such a device construction is not scalable, however, and a large portion of recent efforts in the growth of hBN have therefore centered on epitaxial growth.[5,6,7,8,9,10] Following the success achieved in the growth of graphene, most of these efforts have focused on chemical vapor deposition (CVD) of hBN on metal foils.

Optimization of this growth cannot proceed, however, without a definitive and convenient means of determining the number of monolayers (MLs) present on the surface. Many researchers have attempted to use two methods that are commonly used in graphene systems: optical microscopy, and Raman spectroscopy. Unfortunately, neither method can confidently perform such a measurement in hBN systems.[11,12] Cross-sectional TEM is certainly definitive, but it is a time consuming procedure that characterizes only very small portions of the surface.

[*] feenstra@cmu.edu

Another instrument, the low-energy electron microscope (LEEM), has been established to be a powerful tool in determining the number of MLs of graphene or hBN present on a substrate, in particular by study of low-energy electron reflectivity (LEER) spectra.[10,13,14,15,16] For the case of graphene in particular, it was demonstrated by Hibino and co-workers that by counting the number of *minima* in the 0 – 7 eV energy range of a LEER spectrum, one can confidently determine the number of graphene MLs present.[13] While a qualitative explanation of this phenomenological method was presented in that early work, a quantitative explanation has become available more recently, in the form of a first-principles simulation method for LEER spectra.[17,18]

Here, we present data on single- and multi-ML hBN films grown on polycrystalline nickel foils. These studies were the result of both a desire to characterize the growth of such films for their incorporation into devices, as well as to explore the applicability of quantitative characterization of the hBN based on its LEER spectra. We find that the oscillations seen in the low-energy portion (0 – 7 eV) of the experimental spectra, similar to those seen in graphene, are reproduced by our first-principles method. The participation of various bands of the hBN in the reflectivity, which turns out to be different than for graphene films, is also revealed. Additionally, we find that different surfaces of the Ni yield slightly different LEER spectra, with one source of this difference believed to be Shockley-type surface states of the Ni surface affecting the reflectivity spectra at very low energies.

Compared to prior works involving LEER of two-dimensional (2D) layered materials, the present work achieves greater understanding of how the various bands of the 2D overlayer and the substrate contribute to the LEER spectra. Our experimental results are in good agreement with the prior LEER spectra of Hibino and co-workers for single- and few-layer graphene and hBN films.[10,13] Additionally, our work demonstrates good agreement of theoretically computed spectra with experimental results (focusing on 1 and 2 ML thick films). We find considerable differences between LEER spectra of graphene and hBN, which we demonstrate is due to the different character (symmetry) of their respective electronic bands. Certain bands contribute to the spectra of hBN, but not graphene. It turns out that these same bands for hBN are strongly influenced by inelastic effects (since they are localized on the atomic planes, as opposed to in the interlayer spaces), so a complete computation is needed to understand their contribution to the spectra. We thus understand how the LEER spectra can be used as a means of *chemical identification* between graphene and hBN, and future computations for other materials may yield similar (predictive) capability.

We note that our method of including inelastic effects is more approximate than the prior theory of Krasovskii and co-workers.[19,20,21] Both theories include an energy-dependent imaginary term in the potential, but for the theory of Krasovskii et al. this term is employed within a full solution to the Schrödinger equation whereas in our theory it is applied in a more phenomenological manner, *ex post facto* to an elastic-only solution of the Schrödinger equation. Our theory is thus considerably simplified, since it requires only a standard electronic-structure computation e.g. using the Vienna Ab Initio Simulation Package (VASP), followed by relatively straightforward post-processing of those results. Importantly, we find good agreement between our computed spectra and the experimental ones, thus providing some degree of confidence in the method we employ for including inelastic effects.

## II. Experimental and theoretical methods
### A. Growth

Samples were prepared using two custom-built growth systems. The first is a UHV system with a ~ $1x10^{-10}$ Torr base pressure. This system uses gaseous ammonia and diborane precursors, and was used to prepare the first sample (Sample 1) discussed in this paper. The second system is a low-pressure CVD tube furnace which uses a solid ammonia-borane precursor, and was used to prepare the second and third samples (Samples 2 and 3). The substrates used are 99.9999% pure Ni foils, with thickness of 12.5, 25, 50, or 150 µm. We found that the UHV system allows for the growth of well-ordered hBN multi-layers. Sample 1 was grown on a 150-µm-thick foil in this system and yielded a hBN film with average thickness > 2 ML. In this system, a typical growth starts with the system pumped to the base pressure (~$10^{-10}$ Torr). Then the samples are exposed to a flow of $H_2$ (2 sccms) while heating the chamber and annealing the samples at the growth temperature (~1050°C). Finally, a flow of 2 sccms of diborane and ammonia for is used for the hBN growth (15 min growth time for Sample 1). After the growth is concluded, the samples are cooled down to room temperature in the presence of $H_2$.

Samples 2 and 3 were synthesized in a home built LPCVD system working with ammonia-borane and $H_2$. In this case, 25 micron foils were folded into Ni enclosures[22] were used (the full details of the growth will be published shortly).[23] Briefly, the system is pumped down to the base pressure of ~10 mTorr. Then a flow of 2 sccms of $H_2$ is added and the sample is heated to the growth temperature (1050°C). The samples are annealed for ~30 mins and then exposed to ammonia-borane (which is heated in a separate heater to ~100°C). The growth for samples 2 and 3 were about 10 and 15 min, respectively, yielding hBN films that were thinner than that of Sample 1.

### B. Low-energy electron microscopy

Samples were transferred through air to an Elmitec III system. In this instrument, the sample and electron emitter are kept at high voltage. After leaving the emitter, electrons are accelerated to high-energies (~ 20 keV) into the illumination column wherein the beam is focused. The beam is subsequently diverted towards the sample at normal incidence by a magnetic deflector/beam-separator. Before coming into contact with the sample, the beam may or may not be collimated by an *illumination aperture*. This aperture restricts the area of the beam to sizes between ~ 1 – 7 µm in diameter. Such a confined beam makes possible the performance of selected-area low-energy electron diffraction (µLEED).

Upon approaching the sample, the electrons are decelerated to low-energy (typically 0 – 20 eV) and are then either reflected (or diffracted) from the surface or are absorbed in the sample. Those electrons which are reflected or diffracted are re-accelerated to high energy, passed into the imaging column by the magnetic beam-separator, and are focused either into a real-space image or into an image of the diffraction pattern of the illuminated portion of the surface. Real-space images of the surfaces presented in this work were all done in bright-field imaging mode. Bright-field images are formed by filtering the beam with a *contrast aperture*, allowing only those electrons which are specularly reflected from the sample to pass through; electrons which obtain a non-zero momentum component parallel to the surface during scattering do not take part in image formation in this mode.

### C. First-principles calculations

Our method for theoretically predicting LEER spectra has been previously described, including the important role of inelastic effects in such spectra.[18,24] In the absence of inelastic effects, then for a given structure of the hBN/Ni system we perform a parameter-free computation of the LEER spectra, employing post-processing of electronic states obtained from VASP using the Perdew-Burke-Ernzerhof generalized gradient approximation (GGA) for the density-functional and a plane-wave energy cutoff of 400 eV.[25,26,27,28] This procedure involves computing the states of a supercell consisting of a hBN/Ni/hBN slab with at least 20 Å of vacuum on both sides, and making a detailed analysis of the states thus obtained with those of bulk Ni of the same orientation. The results depend on the structure through the separation of the hBN and the Ni and through buckling of the hBN. To model inelastic effects, a parameter is introduced: the magnitude of the imaginary part of the potential in the solid, $V_i$ , which determines the degree of electron absorption.[19,20,21] In general, $V_i$ will have some energy dependence, so actually more than one parameter is involved. The dominant absorption mechanism at the low energies considered here occurs due to plasmons in the solid. However, the plasmon energy is typically above 15 eV,[19,20,21] which is at the upper edge of the energies that we consider. For lower energies, the absorption mechanisms are less well understood, and they may involve surface defects or disorder.[19]

In our prior work,[24] considering single- or few-layer graphene or hBN on various substrates, we have argued that a linear form for $V_i$ given by 0.4 eV + 0.06 $E$ where E is the energy relative to the vacuum level fits the experimental reflectivity data fairly well (for energies below the plasmon turn-on), and we continue to use this form for all of our analysis in the present work. Additional parameters in the inelastic analysis are an "inner potential" for the electrons in the solid, and a "turning point" (relative to the surface plane) for the incident electrons. We have demonstrated previously that our results are very insensitive to the choice of these parameters,[24] and we employ values of 13 eV and 0.9 Å, respectively, for all of our analysis.

In the theoretical results shown throughout this work we will compare the computed reflectivity with the band structure of the hBN overlayer (i.e. following the useful presentation method introduced by Hibino and co-workers[10,13]) and of the Ni substrate. To align these respective band structures with the reflectivity spectra, we precisely align the respective potentials from VASP computations for bulk hBN or bulk Ni with that for the hBN/Ni slab computation, focusing in all cases on alignment of the potential minima that occur at the locations of atomic planes.

## III. Experimental Results

Figure 1 shows a series of four LEEM images of a multilayer hBN island from Sample 1, illustrating how the contrast for this island containing different numbers of MLs evolves as a function of energy. LEER spectra of points indicated in the images are shown in Fig. 1(e), labeled according to the number of hBN MLs we interpret as being present at that location, as discussed below in Section IV. The spectra of Fig. 1 are plotted as a function of the voltage difference between the sample and the electron emitter in the LEEM. This difference, minus the work function difference between sample and emitter (together with a small term arising from the width of the emitted electron distributions), equals to the electron energy relative to the vacuum level of the sample.[29] The work function difference between sample and emitter is typically 1 – 2 eV, varying with the precise location on a given surface (and also from sample to sample). Hence, the *mirror-mode onset* of the reflectivity spectra, i.e. the rapid rise in the reflectivity that occurs as a function of decreasing voltages, is typically seen at 1 – 2 V sample voltage. We have developed a detailed fitting procedure to extract this work function difference from the measured spectra;[29] utilizing that method, in Section IV we show the spectra plotted as a function of electron energy. Examining the mirror-mode onsets in Fig. 1, it is apparent that for some spectra (e.g. D and E) there is a maximum in the reflectivity just below the onset (i.e. at about 1 V). The apparent reflectivity is actually greater than unity at these maxima, which is an artifact of the measurement due to electrons being transferred into the measurement area by lateral electric fields arising from work function variations on the surface (such artifacts are greatly reduced at energies above the mirror-mode onset, due to the higher electron energies there).[29]

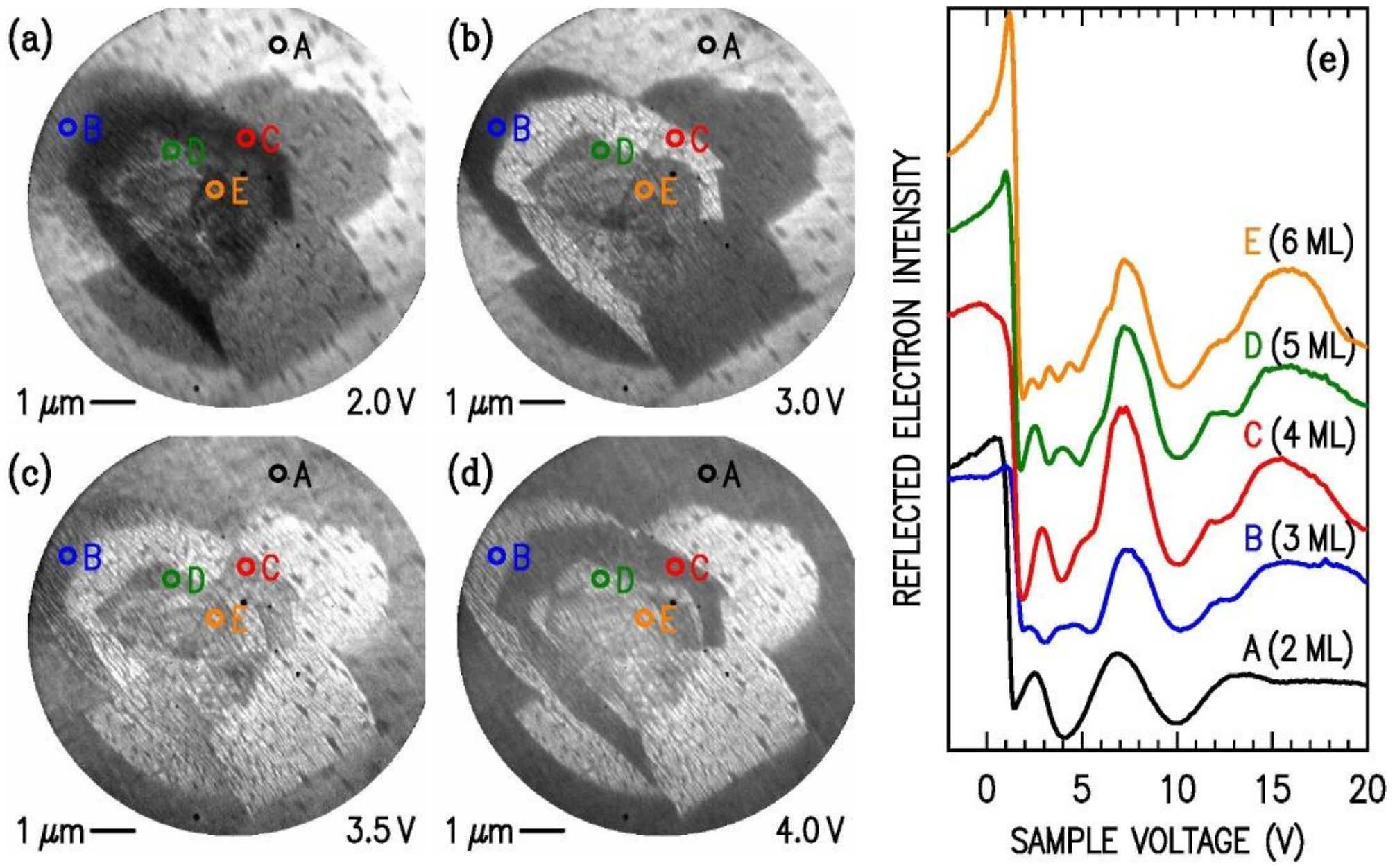

FIG 1. (a) – (d) LEEM images from Sample 1 showing a multilayer hBN island at different sample voltages. (e) LEER spectra extracted from the locations indicated in the images and labeled according to the number of hBN MLs present.

Generally speaking, the features which are of greatest importance in LEER spectra are the location of reflectivity *minima*. For multilayer two-dimensional materials, such minima oftentimes can be interpreted as arising from *interlayer states*, i.e. special electronic states that exist in the interlayer spaces between the atomic planes.[13,17,18] When the energy of an incident electron is coincident with one of these states, then the electron can easily couple (connect with) the interlayer state and be transmitted into the material. Hence, a local minimum in the reflectivity is produced. The minima seen in Fig. 1 located in the oscillatory portion of the spectra, from ~ 2 – 7 V, arise from such interlayer states. The minimum located at ~ 10 V also arises in part from such states, but the mechanism in this case is more complicated as will be discussed in Section IV. (This latter minimum is quite important because it does *not* occur in graphene, hence providing chemical identification of hBN relative to graphene for situations where the two materials need to be distinguished).[29,30]

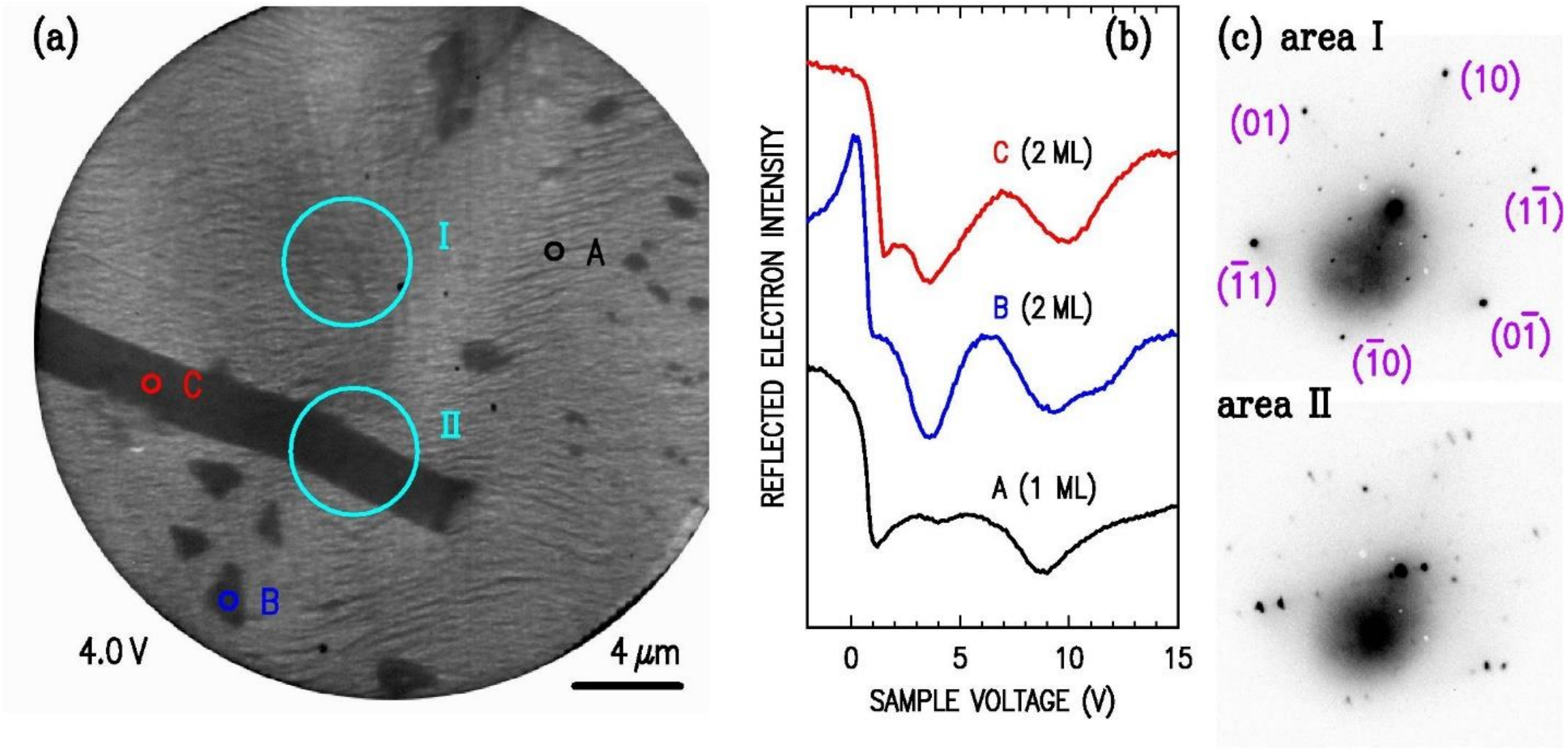

FIG 2. (a) LEEM image from Sample 2 at the specified sample voltage. (b) LEER spectra from surface locations indicated in (a). (c) μLEED patterns shown in reverse contrast, acquired from areas I and II indicated in (a), with electron energy of 45 eV.

Figure 2 shows results obtained from Sample 2, for which the hBN coverage is found to be lower than than that of Sample 1. The majority of the surface shown in Fig. 2 is covered with what we interpret as single ML hBN, i.e. with a single dominant reflectivity minimum near 9 V as in spectrum A. This spectrum also has a weak local minimum near 4 V, which is replicated in the theory of Section IV and is found to be *not* related to the interlayer state at 3 – 4 V seen in the 2-ML LEER spectra, B and C. These latter 2-ML spectra appear at isolated areas on the surface. They have two distinct reflectivity minima, the one at 3 – 4 V and another at 9 – 10 V. Importantly, the higher voltage minimum sometimes appears with a shoulder located at ~ 11 V, as in spectrum B, and other times does not have this shoulder, as in C (the latter case closely resembles the 2-ML spectrum A of Fig. 1). Associated with this presence or absence of the shoulder, the spectra also reproducibly show different behavior at low voltages near the mirror-mode onset at ~ 1 V. For spectrum C in Fig. 2, there is a distinct downturn in the reflectivity as the onset is approached as a function of decreasing voltage. In contrast, spectrum B does not show this downturn, but rather, there is only a flattening of the reflectivity before the sharp increase of the mirror-mode onset. Again, these differences between spectra B and C are reproducibly seen for all spectra that we have acquired from 2-ML thick hBN films.

Regarding the origin of these differences between spectra B and C of Fig. 2, we note that the distinct rectangular shape of the 2-ML thick area from which spectrum C is acquired is suggestive that possibly this surface area terminates a different grain of the Ni substrate compared to that of the surrounding area (i.e. the surrounding area including the more triangular-shaped 2-ML islands such as the one from which spectrum B is acquired). In an effort to identify such grains, in Fig. 2(c) we present μLEED patterns from specific surface areas. Note that in these and all other μLEED patterns presented in this work, the large, intense features located just below and to the left of the center of the patterns are a result of secondary electrons, i.e. electrons which undergo *inelastic* scattering, and hence should be disregarded in any analysis of the patterns. For the case of area I of Fig. 2(c), we see that the most intense spots (aside from the (00) spot at the center) are located towards the edge of the pattern: six spots in a hexagonal arrangement, labeled (10), (01), ..., in the pattern, with wavevector consistent with that of hBN or Ni(111) (only an 0.5% difference in their lattice constants). We associate these spots with the hBN, since we know from the LEER spectra that hBN is covering the surface. We observe only the single set of 6 primary spots at this wavevector, i.e. no other spots with rotated alignment, and also we do not see any spots elsewhere in the pattern that do not have hexagonal symmetry. We therefore tentatively conclude that this area is 1 ML hBN on Ni(111), and that the hBN spots have eclipsed the spots from the Ni. Comparing this to the pattern of area II in Fig. 2(c), we see the latter exhibits many more spots, with some higher order spots running along parallel lines. While we cannot determine the surface orientation of the underlying rectangular-shaped Ni grain, it appears not to have (111) orientation, and we believe it likely is either (100) or (110) oriented, i.e. due to its rectangular shape.

More definitive identification of grain orientations is provided in data from Sample 3, shown in Fig. 3. The LEEM images of Figs. 3(a) and (b) clearly show the presence of different regions of the surface, which again we interpret as different grains of the Ni substrate. Boundaries between these grains are indicated in Fig. 3(a). LEER spectra B – D of Fig. 3(c), acquired from 1-ML and 2-ML areas of hBN, are in very good agreement with those of Figs. 1 and 2. An additional spectrum, curve A which arises from the dark patch of Fig. 3(b), is quite different than any that we have reported thus far. This spectrum is characterized by a lack of distinct reflectivity minimum, instead gradually falling until it becomes flat and featureless above 10 V. We attribute this spectrum to a bare Ni surface, based on the similarity of its spectrum to that we have seen on other bare metal surfaces.[18,31]

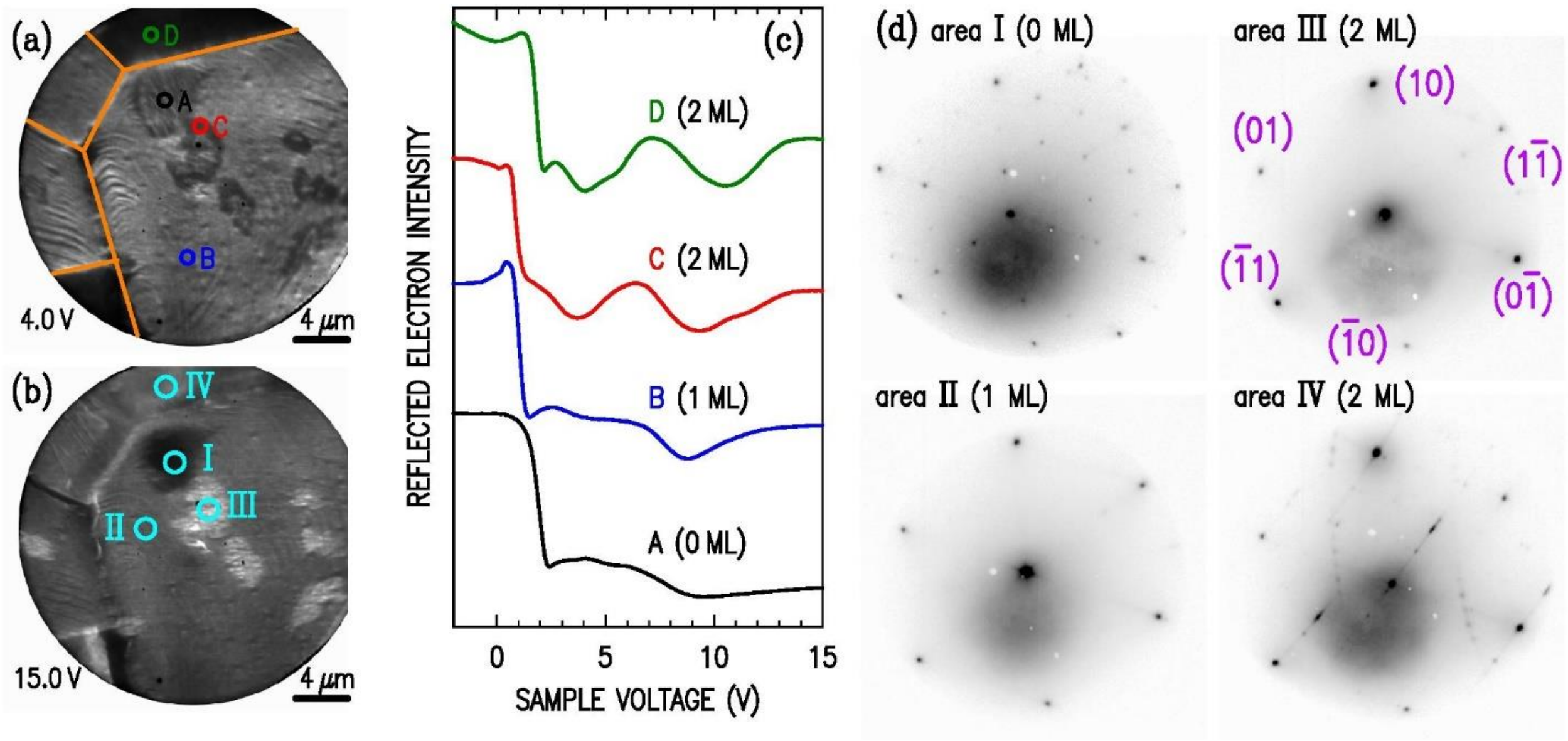

FIG 3. (a,b) LEEM images of Sample 3 at two different sample voltages. Grain boundaries of the underlying Ni substrate are emphasized in (a) with orange lines. (c) LEER spectra of the points indicated in (a). (d) μLEED patterns from locations I – IV indicated in (b), acquired at energies of 45, 40, 40, and 45 eV, respectively.

The associated diffraction pattern, area I of Fig. 3(d), shows a hexagonal arrangement of {10} spots about a (00) spot. Additional, weaker spots are also seen, presumably arising from some reconstruction and/or adsorbates of the Ni surface. Areas II and III display six dominant {10} spots, arising presumably from the hBN, with the former area covered by 1 ML of hBN and the latter by 2-ML-thick islands. Finally, examining the pattern of area IV, which we have interpreted as arising from a different grain orientation of the Ni substrate, we again see six {10} spots arising from the hBN. However, this pattern also displays exhibits many more spots, with some higher order spots running along parallel line, similar to the pattern of area II of Fig. 2 and typical of a vicinal surface.[32] Hence, from this diffraction pattern, we can be confident that this grain of the underlying Ni substrate has a surface orientation that is quite different from (111).

Incidentally associated with this change of Ni surface orientation, we also point out the distinct striations seen in the LEEM images of the surface with (111)-orientation (or nearly that orientation), i.e. the wavy lines separated by tenths of a μm and with varying bright/dark contrast, seen almost everywhere in Fig. 2(a) and in some surface areas of Figs. 3(a) and 3(b). We interpret these striations as arising from step bunches formed by faceting of the surface,[23] i.e. due to the fact that the surfaces are *vicinal* with orientation that is close to, but not exactly, along a (111) direction. Such faceting is commonly observed e.g. for graphene covering Cu(100) vicinal surfaces,[16] but for the present case we observe the striations only on the Ni grains with (111) vicinal orientation, and *not* for the grains with surface orientation much different than (111). (In principle, the striations observed in the LEEM images might arise from individual steps, rather than step bunches, but the common occurrence of step bunches in prior work makes it likely that such bunches occur for the present samples as well).[16,18,29]

One additional feature of the hBN diffraction patterns to note is that they have 3-fold symmetry, i.e. with the (10), $(\bar{1}1)$, and $(0\bar{1})$ spots have intensity that differs significantly from that of the (01), $(\bar{1}0)$, and $(1\bar{1})$ spots. This asymmetry is seen most clearly in the patterns of areas II and IV of Fig. 3, but it is invariably present on all hBN-covered areas that we have studied (LEED patterns are generally acquired with 5-eV spacing from 40 to 90 eV, and the asymmetry is clearly apparent in the majority of the patterns). Such intensity asymmetry is expected from the 3-fold structure of the hBN.[33] We also find that situations such as in Fig. 3 when the surface contains 2-ML-thick islands surrounded by predominantly 1-ML thick hBN, that the 3-fold hBN spots have the same orientation (same set of three {10} spots being the most intense) irrespective to whether the pattern is acquired from 1-ML or 2-ML thick areas. This result indicates that the top-most layer of each surface area is oriented in the same way, implying that any additional layers have actually grown *underneath* the first layer. This conclusion is the same as that reported in recent study of graphene growth on Cu substrates.[16] Additional data from Sample 3, presented elsewhere, reveals an unchanging orientation of the diffraction spots even for μLEED patterns that were acquired at surface locations separated from each other by 100's of μm, indicative of very large single-crystal domains of the hBN.[23]

## IV. Theoretical Results

To validate the identifications made in the previous Section concerning the respective numbers of hBN layers in the LEEM images, it is necessary to compute the LEER spectra and to demonstrate agreement between the theoretical and experimental spectra. As a byproduct of that comparison, we are also able to obtain some structural information, such as the separation between the Ni surface and the adjacent BN layer, as well as the buckling (difference between B and N heights) of that hBN layer. To date, LEER spectra of hBN have not been modeled in detail. However, for the case of hBN on Co(0001), Orofeo et al. have presented experimental LEER spectra, and they have qualitatively argued how those spectra can be interpreted based on the band structure of the hBN.[10] We have previously presented theoretical simulation of such spectra, including the influence of inelastic effects which were shown to be relatively strong.[24] (We also discussed the influence of oxidation of the surface in that work, although we realize now that oxidation is not a requirement for obtaining agreement between experimental and theoretical spectra, a point that we return to in Section V).

There are four bands of hBN located in the range 5 – 17 eV above the Fermi energy, $E_F$, that contribute to the low-energy LEER spectra. These bands are pictured in Fig. 4(a), focusing on wavevector values between Γ to A which are the only states relevant to LEER (the full DFT-computed band structure has been presented previously,[34] and is in agreement with prior work[35]). Four analogous bands exist for graphene.[13,17] The lowest energy band for both hBN and graphene has character that is dominantly composed of *interlayer states*, i.e. plane-wave type states that exist predominantly in the spaces between the 2D layers. These interlayer states give rise to pronounced minima in the reflectivity spectra. The higher lying bands have character that can be directly interpreted in terms of linear combination of atomic orbitals of the atoms.[31] For graphene, this interlayer band is not coupled to the three higher lying bands due to the inherently different symmetries of the respective states,[30] so those higher bands make no contribution to the reflectivity spectra. However, for hBN, the higher bands are coupled to the lower interlayer band (the symmetry of the states is lower than for graphene, i.e. due to the different character of the atomic orbitals of B compared to N), and hence they make a substantial contribution to the spectra. The contribution of the higher bands to the spectra was demonstrated experimentally by Orofeo et al.[10] and theoretically in our prior work;[24] this coupling between the hBN bands plays an important role in the results presented below. Also seen in Fig. 4 are additional bands with energies > 20 eV. Some of these bands, the lowest energy one in particular, also have interlayer character and give rise to minima in the reflectivity spectra. An analogous set of high-energy band(s) with interlayer character exists for graphene.[24,36,37]

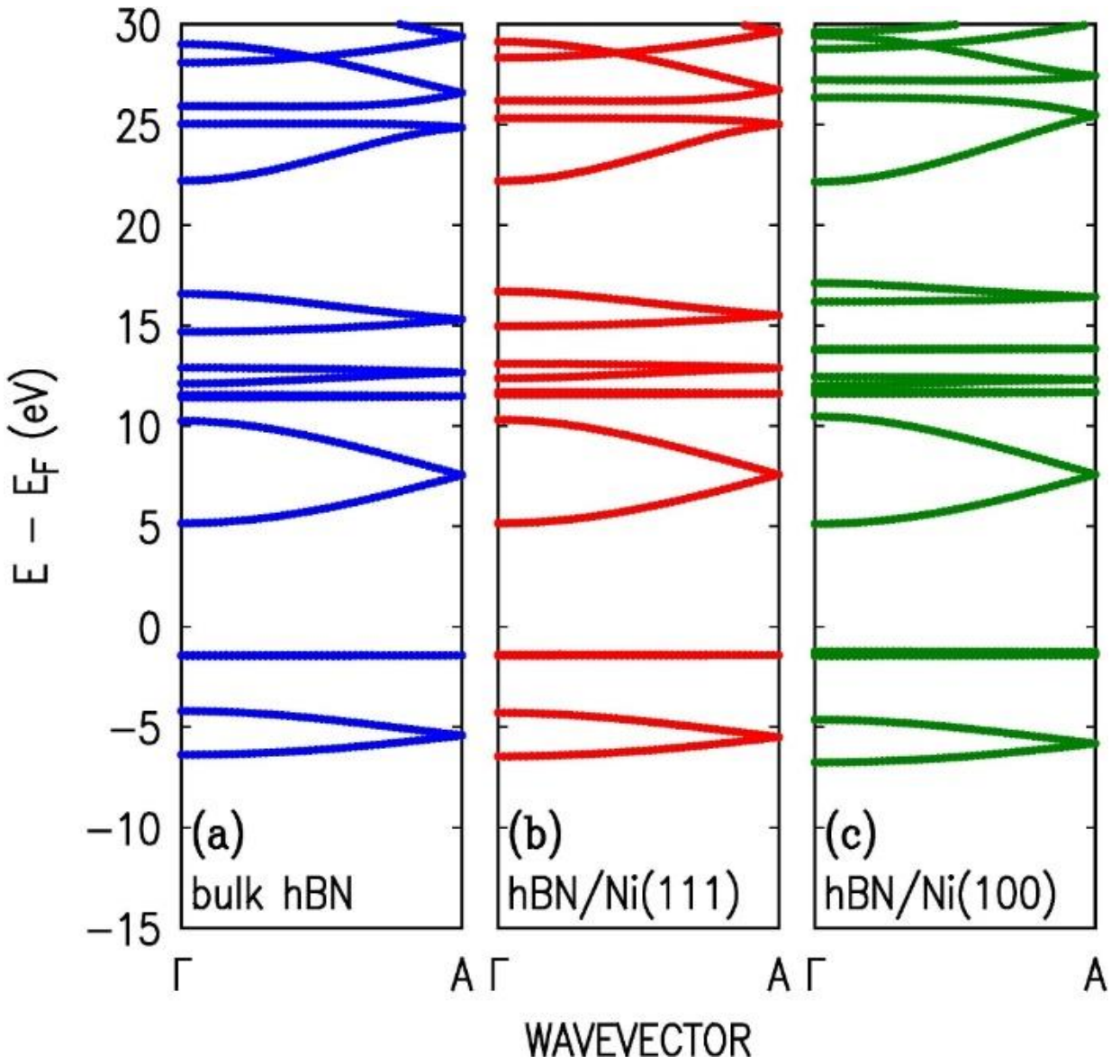

FIG 4. Computed band structures in the Γ-A direction for (a) bulk hBN, (b) hBN strained by $-0.5\%$ (compressive) to match Ni(111), (c) hBN strained by $-0.5\%$ and $-4.3\%$ in two orthogonal surface directions, to match Ni(100). Energies are shown relative to the Fermi energy, $E_F$.

Regarding the epitaxial fit between the hBN and the nickel, we employ a perfect, 1×1 fit for the Ni(111) surface, corresponding to $-0.5\%$ strain (compressive) of the hBN. For modeling of a non-(111) orientation we employ the Ni(100) surface, with a 5×1 fit for a 6×1 hBN unit cell (as previously observed for hBN on Cu(100) by Liu et al.),[33] corresponding to $-0.5\%$ and $-4.3\%$ strain for the hBN in the two orthogonal surface directions. Figures 4(b) and 4(c) shows the computed band structures for bulk hBN, given these strain values. The results for the (111)-strained case, Fig. 4(b), are only slightly different than those for the unstrained, bulk hBN as in Fig. 4(a). Similarly, results for the (100)-5×1 cell, Fig. 4(c), are also fairly close to the unstrained case. In contrast, if we use a 2×1 cell for the (100) surface we obtain hBN bands that differ from the unstrained case by many eV, producing unphysical reflectivity spectra. In our plots below of the theoretical and experimental LEER spectra we will, for simplicity, compare the energetic locations of their features with those of the predicted bands of the unstrained hBN (i.e. Fig. 4(a)).

Figure 5(a) shows the computed LEER spectra for 1 ML hBN on Ni(111), with energies now plotted relative to the vacuum level for the system. No structural relaxation of the atoms is included in the computation, with the Ni-Ni spacing taken to be that of bulk Ni and the BN-Ni spacing being a parameter in the computations. Prior experimental and theoretical work indicates a relatively strong interaction between the BN and the Ni, with an average BN-Ni separation in the range 2.0 – 2.2 Å and a BN buckling (difference between B and N vertical heights) of 0.07 – 0.2 Å with the B atom closer than the N to the Ni surface plane.[38,39,40] In Fig. 5 we use an average BN-Ni separation of 2.08 Å and buckling of 0.2 Å, and we show results both with and without inelastic effects. Following prior work we have assumed a lateral registration of the hBN and Ni with the N atoms on top of surface Ni atoms,[38,39,40] although we obtain essentially identical reflectivity results (i.e. to within the size of data points used for plotting) when the B atoms are placed atop the Ni. Figures 5(b) and 5(c), respectively, show the band structure in the direction perpendicular to the surface for the underlying Ni substrate and for a hypothetical, bulk layer of hBN on the surface. In fact, the buckling of the hBN would produce significant changes to this band structure, but nevertheless for qualitative purposes it is useful to display this unbuckled bulk band structure (i.e. following the approach of Orofeo et al.[10]). The energy alignment for these bands is determined by comparing the potentials of the bulk bands with those of the hBN on Ni computation.

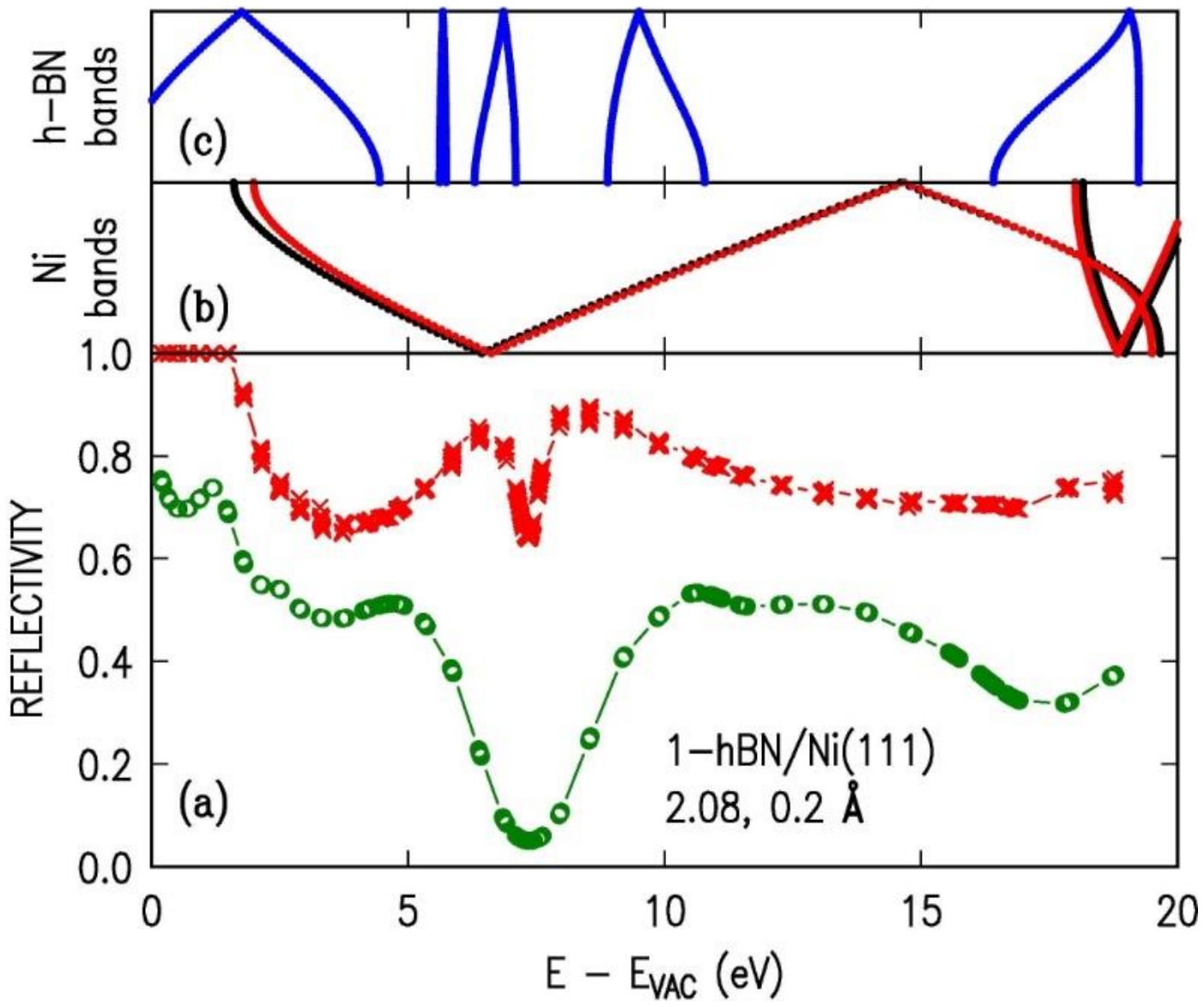

FIG 5. (a) Computed reflectivity spectra for 1 ML hBN on Ni(111), with (circles) and without (x-marks) inelastic effects, averaged over minority and majority spins. Average BN-Ni separation is 2.08 Å and buckling is 0.2 Å. (b) Bulk Ni band structures in (111) direction (majority spin band has noticeably lower energies than the minority spin band, for energies < 5 eV). (c) Bulk hBN band structure in (0001) direction.

The interpretation of reflectivity spectra in the low-energy range involves associating *minima* in the spectra with transmission resonances arising from interlayer states, as well as consideration of possible band structure effects associated with the overlayer or the substrate.[13,17,18,24] Let us first consider the spectrum of Fig. 5 that neglects inelastic effects. A reflectivity of unity is obtained for energies below 1.6 eV, associated with the onset of the Ni majority-spin nearly-free-electron (NFE) band at that energy. For higher energies a reduced reflectivity is found, arising from the hBN band with strong interlayer character seen at 0 – 5 eV in the hBN band structure. However, for the BN-Ni separation of 2.08 Å the interlayer space is too small to support a well-defined interlayer state; hence, this reflectivity minimum near 3.5 eV is rather broad. As discussed above, the higher hBN bands can couple to the interlayer band and produce their own reflectivity minima. This does indeed occur, as seen by the distinct minimum at 7.3 eV relative to the vacuum level. Inclusion of inelastic effects causes this minimum to become more pronounced, and it turns out to dominate the spectrum. At higher energies, near 17 eV, a smaller reflectivity minimum is seen; it arises from higher lying bands that have interlayer character (one of which is seen at the upper end of the hBN band structure in Fig. 5).

The dependence of the computed reflectivity spectra on BN-Ni separation and buckling is shown in Fig. 6. The prominent minimum in the reflectivity near 7 – 8 eV persists for all values of BN-Ni separation and for all buckling values ≥ 0. Positive buckling (i.e. B atoms closer than N to Ni plane) values cause this minimum to deepen slightly and to shift down in energy. Negative values of buckling (N atoms closer to Ni plane) cause the minimum to become more shallow and disappear, being replaced by a lower energy minimum near 5 eV. Comparing to the experimental data at the bottom of Fig. 6, we can confidently rule out negative buckling values as being inconsistent with experiment. Furthermore, for zero buckling (left-hand panel of Fig. 6(a)), the shapes of the computed reflectivity minima near 7 eV do not provide a good match to the experiment. Visual comparison of the features of the experimental and theoretical spectra, including the minima at 7 – 8 eV, the small dip near 3 eV, and the overall shapes near the low- and high-energy ends of the spectra, qualitatively suggests a best fit between the two for BN-Ni separations of 1.9 – 2.1 Å and a buckling of 0.1 – 0.2 Å (although an overall reduction in intensity of all spectral features is seen in the experimental spectrum compared to the theoretical ones). These results for the structural parameters fall well within the range of previously determined values.[38,39,40]

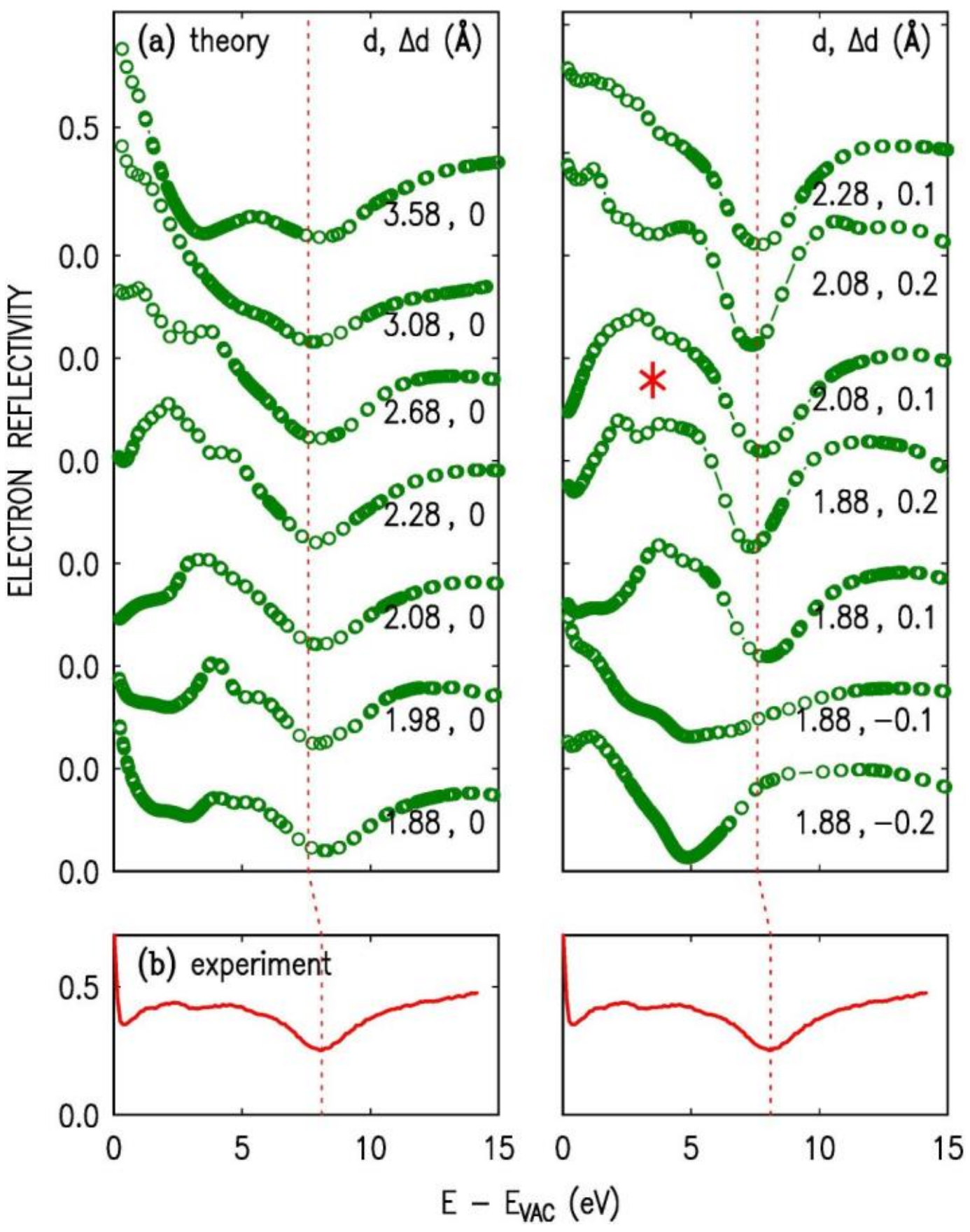

FIG 6. (a) Computed reflectivity spectra (averaged over spins) for 1 ML of hBN on Ni(111), as a function of d (the average BN-Ni vertical separation) and Δd (BN vertical buckling). N atoms are atop Ni, and positive buckling refers to a smaller B-Ni vertical separation than that of N-Ni. (b) Experimental reflectivity spectrum. The *same* experimental curve is repeated in the right- and left-hand panels, for the purpose of comparison with theory. Dotted lines display estimated shifts between theory and experiment (see text). Theoretical curves with the best match to experiment are indicated by an asterisk.

Figure 7 shows the computed LEER spectra for 2 ML hBN (with AA' stacking) on Ni(111), using an average separation between the top Ni layer and the adjacent BN layer of 2.03 Å and buckling of 0.1 Å for that BN layer. Zero bulking of the second BN layer is assumed. The separation between the two hBN layers is assumed to be equal to the value for bulk hBN, 3.30 Å from Ref. [41], with this value taken to be the separation between the second BN layer and the outermost layer of B or N in the first BN layer. The prominent minimum seen at 3.0 eV in the spectrum arises from an interlayer state localized between the two hBN layers; it derives from the lowest hBN band. Two higher energy minima are seen in the spectrum that neglects inelastic effects, at 7.2 and 9.2 eV; these derive from high hBN bands that are coupled to the lower one, and these minima evolve into a broad, asymmetric minimum when inelastic effects are included. As discussed above in connection with Fig. 4, the nature of the states that compose the 3-eV minimum compared to the one at 7 – 10 eV are quite different; examination of the VASP pseudo-wavefunctions of these states (shown in Supplementary Material) reveals that the states associated with the interlayer band near 3 eV are concentrated *between* the hBN layers, whereas those for the 7 – 10 eV minimum are concentrated *on* the hBN planes. Additional minima are seen at even higher energies, 17.6 and 19.5 eV, in the spectrum that neglects inelastic effects; these minima also broaden considerably when inelastic effects are included.

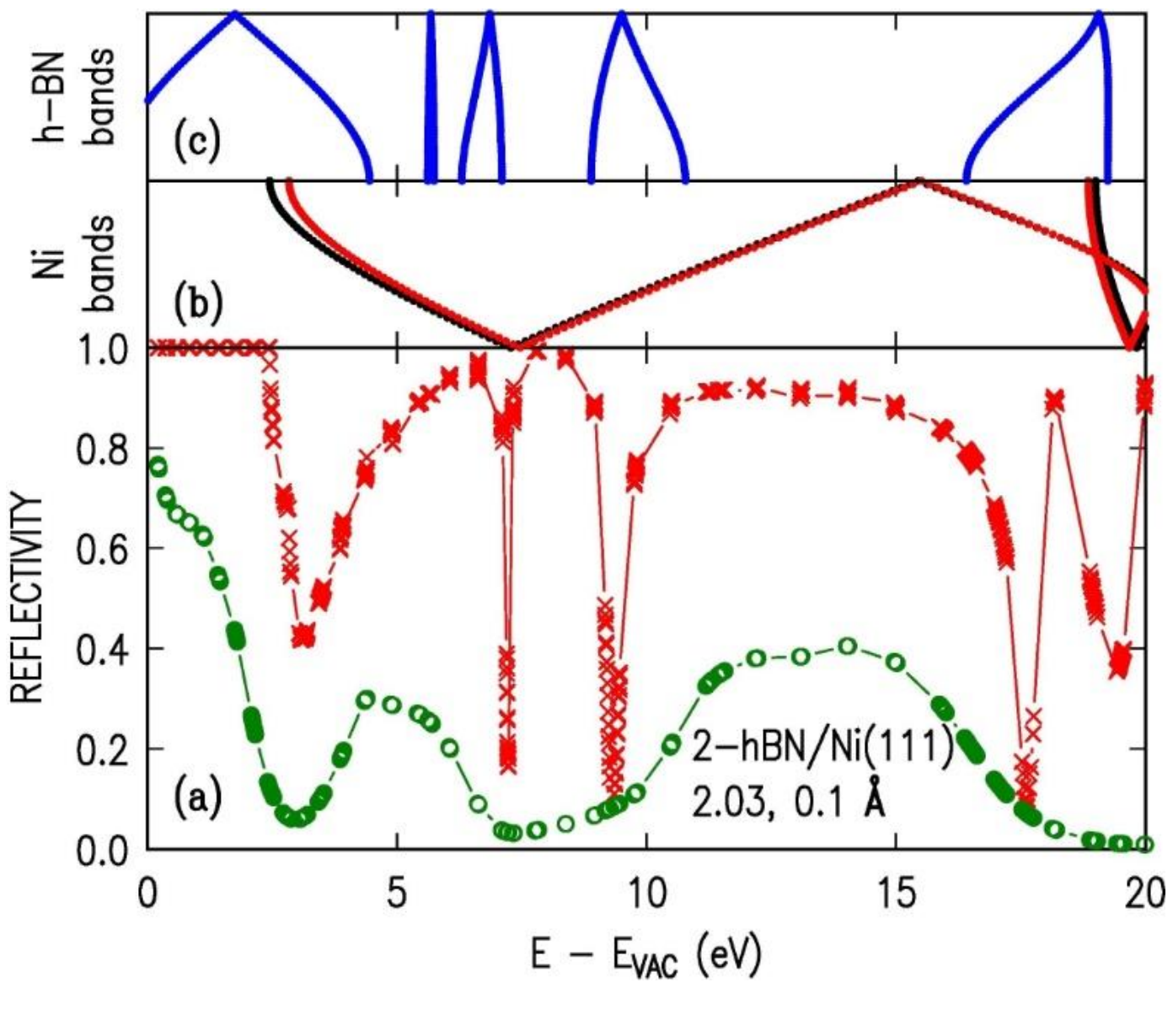

FIG 7. Same caption as Fig. 5, but for 2 ML of hBN on Ni(111), with average separation between Ni and the adjacent BN layer of 2.03 Å, and 0.1 Å buckling of that BN layer (B atoms closer than N to the Ni plane).

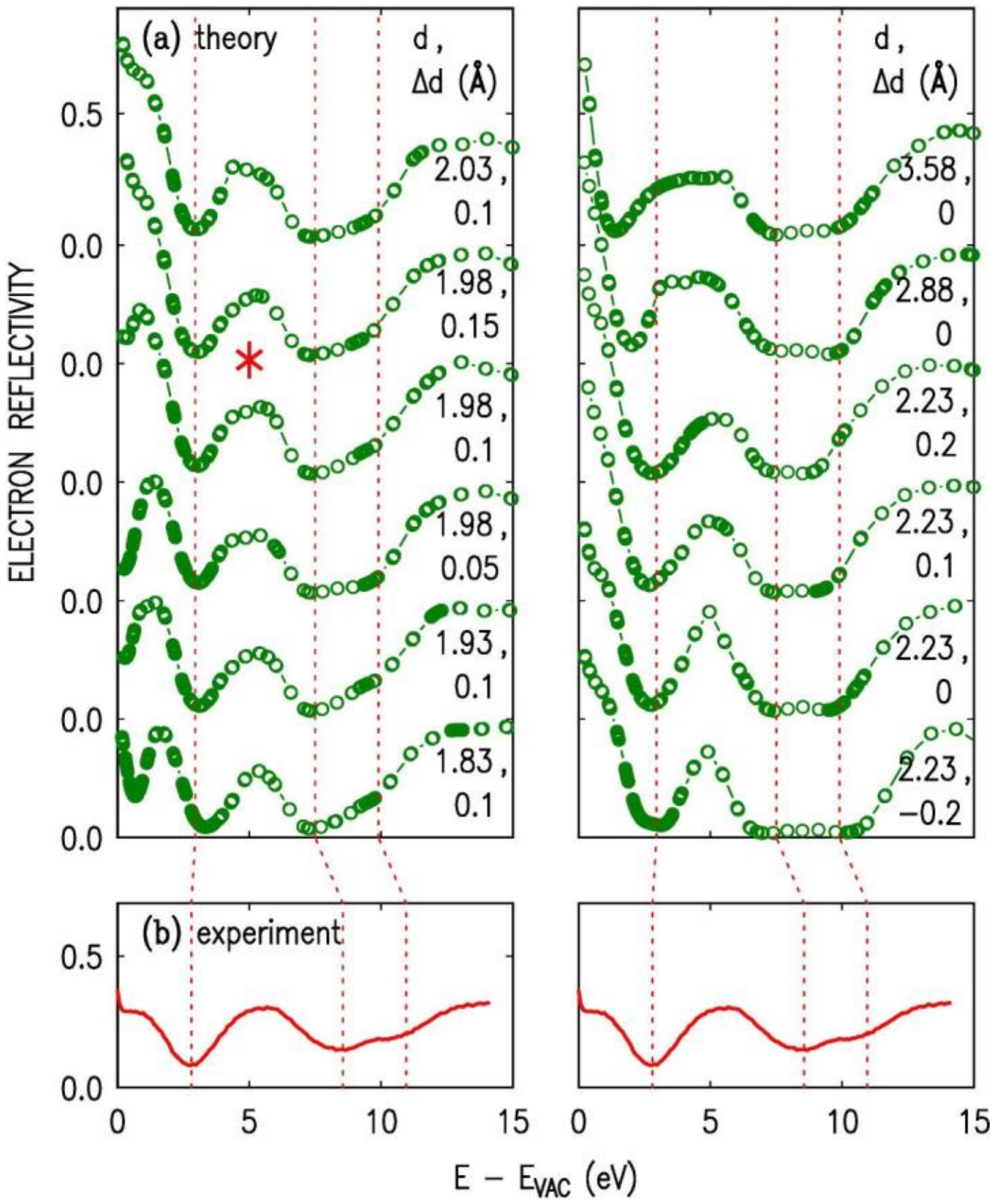

FIG 8. (a) Computed reflectivity spectra (averaged over spins) for 2 ML hBN on Ni(111), as a function of d (average BN-Ni vertical separation) and Δd (buckling). N atoms are atop Ni, and positive buckling refers to a smaller B-Ni vertical separation than that of N-Ni. (b) Experimental reflectivity spectrum. The *same* experimental curve is repeated in the right- and left-hand panels, for the purpose of comparison with theory. Dotted lines display estimated shifts between theory and experiment (see text). Theoretical curves with the best match to experiment are indicated by an asterisk.

The dependence of the 2-ML reflectivity curves on the BN-Ni separation and buckling is shown in Fig. 8. The asymmetry of the broad (double) minimum extending over 7 – 10 eV is seen to depend on both the BN buckling and the BN-Ni separation. Comparing to the experimental data at the bottom of Fig. 8, we see that values near 2.0 and 0.1 Å for these parameters are certainly consistent with the data. Larger values of both separation and buckling can be excluded, since then this minimum take on a flat-bottomed appearance (e.g. for the spectra in the right-hand panel of Fig. 8(a)). Smaller values of separation cannot be excluded on the basis of this particular feature in the spectra.

Given the comparison of experimental and theoretical spectra in Figs. 6 and 8, it is apparent that energy shifts can occur between them. In particular, a shift of 0.5 eV was found in Fig. 6 the experimental and best-fit theoretical locations of the reflectivity minimum near 8 eV. For Fig. 8, the broad minimum at 7 – 10 eV in the theory is shifted substantially compared with experiment, by about 1.1 eV. In contrast, for the lowest energy reflectivity minimum near 3 eV in Fig. 8, the shift by only about 0.3 eV (which is not much larger than our uncertainty of ±0.1 eV). One contribution to such errors arises from inaccuracy in our determination of work functions for the surfaces that we are dealing with, i.e. due to the GGA density-functional approximation that we are using, as previously discussed for the case of graphene on metal surfaces.[18] Such errors, however, amount only to a few tenths of an eV typically.[18] In the present case we find significantly larger errors associated with the energies of the states (relative to the vacuum level) that are localized on the atomic planes. Thus, it appears that the inaccuracy of the GGA treatment appears to vary with the character of the state in question (being relatively small for the interlayer states in particular). Future results, employing the GW method, would be useful further explore this issue.

An important aspect of the low energy minimum near 3 eV is apparent if we examine the theoretical spectrum for relatively small BN-Ni separations, i.e. the 1.83 Å value with 0.1 Å buckling displayed in the lower left-hand corner of Fig. 8(a). We see there that a *second* minimum appears to even lower energy, 0.6 eV for this particular spectrum. The presence of this additional minimum is also seen in some of the other theoretical curves by a flattering and/or downturn of the reflectivity as the energy approaches zero. A similar flattening is apparent in the experimental spectrum shown in Fig. 8(b), and is even more pronounced in the non-(111) spectrum of Fig. 9(b). The presence of this additional minimum is difficult to understand within our previously described general interpretation of reflectivity spectra.[18] Specifically, we argued that for two layers of a 2D material on a substrate we expect a low-energy interlayer state (and hence a reflectivity minimum) associated with the space between the two 2D layers. Additionally, if the separation between the lowest 2D layer and the substrate is sufficiently large separation, $\gtrsim 3$ Å, then a second interlayer state (and second reflectivity minimum) will form. However, the 1.83-Å separation considered in the lower left corner of Fig. 8(a) is much too small to support such a state. In Section IV we further discuss this second, low-energy reflectivity minimum, arguing that it arises from Shockley-type states arising from the band gap at the $\overline{\Gamma}$-point that occurs for Ni surfaces.[42,43,44]

Let us now turn to LEER spectra obtained from hBN on the non-(111) Ni surfaces discussed in Section III, which we model here in terms of the Ni(100) surface. Figure 9 shows results from a set of computations of 2-ML hBN on Ni(100), compared to the experimental spectrum of 2-ML hBN. As seen in the experimental spectrum, the features we associate with this non-(111) surface are the presence of the downturn (with decreasing energy) at low energies, and the lack of the plateau (shoulder) in the reflectivity for energies above the 9 eV minimum. If we first compare these 2 ML computational results to those for Ni(111) in Fig. 8, we find that for the (100) surface the "extra" minimum at very low energy is indeed more pronounced. For example, the spectrum for separation of 2.28 Å and zero bulking in Fig. 9 clearly reveals the extra minimum, at 0.2 eV, whereas in Fig. 8 a comparable spectrum with separation of 2.23 Å and zero buckling does not show it at all. This enhanced presence of the extra low-energy minimum in the (100) computations is consistent with its identification, discussed in Section V, as arising from a Shockley-type surface state for that surface, since the surface band gap is substantially higher lying for the (100) surface compared to the (111).[42,43,44] The more pronounced presence of the extra very-low-energy reflectivity minimum in the (100) computations is also fully consistent with the experimental results of Figs. 8(b) and 9(b).

Comparing the experimental and theoretical spectra of Fig. 9, it appears that the best fit occurs for an average separation of about 2.1 Å, and with BN bulking of about 0.0 – 0.1 Å. The lowest energy partial minimum just above 0 eV is well described in the theory as just discussed, and similarly for the minimum near 2 eV arising predominantly from the interlayer state between the two BN layers. Regarding the higher lying minimum extending over 7 – 10 eV in the theory, it only approximately matches the minimum centered at 8.5 eV in the experiment. The theoretical minimum is too broad, and it displays some small features (ripples) at energies just above 8 eV. In this regard, we note that in our computations we are maintaining identical buckling for all B and N atoms in the first layer (as well as using zero buckling for all B and N atoms in the second layer). However, given the 6×1 BN / 5×1 Ni(100) fit of their unit cells,[33] it seems very likely that any buckling that does occur in the first

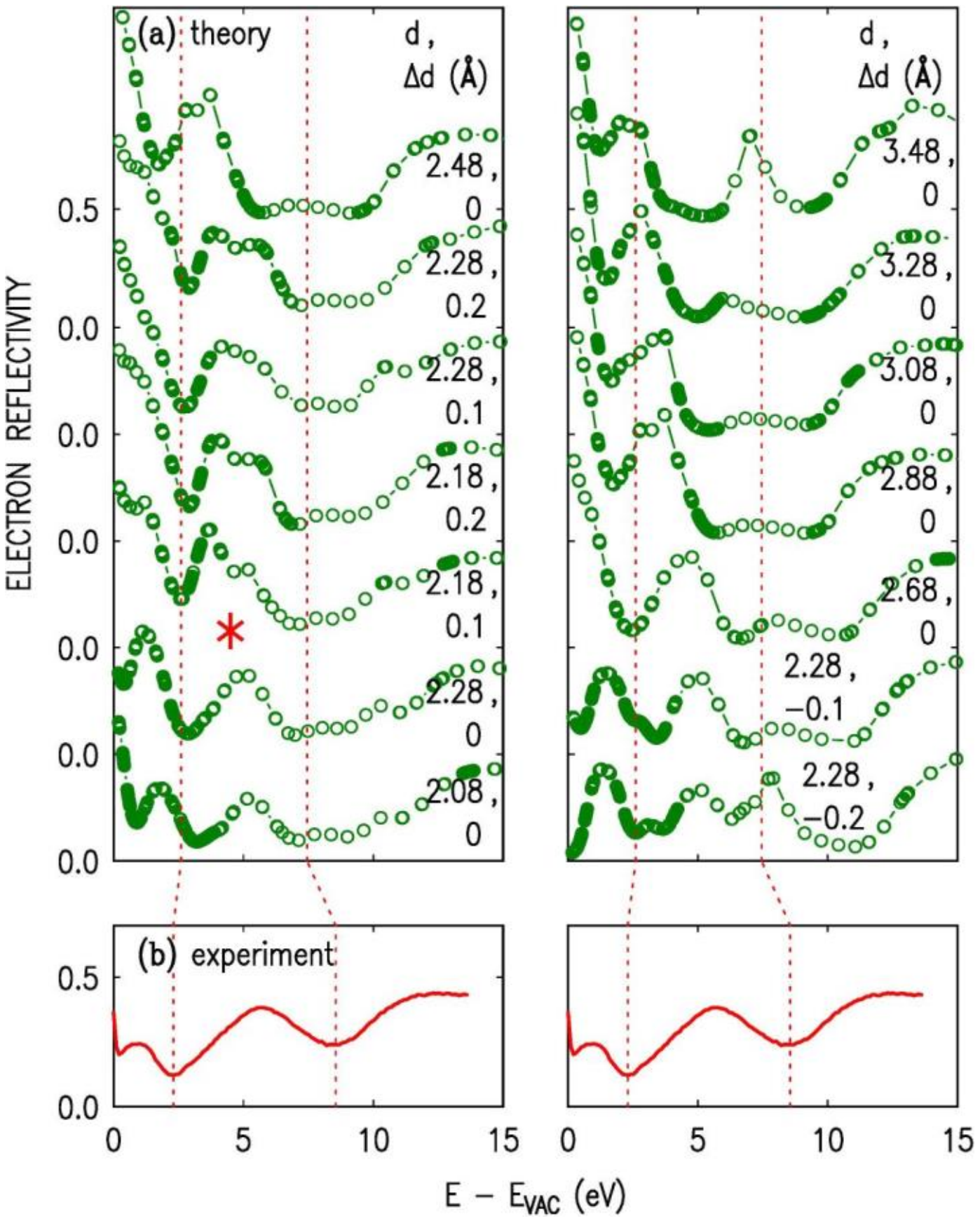

FIG 9. (a) Computed reflectivity spectra (averaged over spins) for 2 ML hBN on Ni(100), as a function of d (average BN-Ni vertical separation) and Δd (buckling). N atoms are atop Ni, and positive buckling refers to a smaller B-Ni vertical separation than that of N-Ni. (b) Experimental reflectivity spectrum. The same experimental curve is repeated in the right- and left-hand panels, for the purpose of comparison with theory. Dotted lines display estimated shifts between theory and experiment (see text). Theoretical curves with the best match to experiment are indicated by an asterisk.

BN layer will varying depending on lateral location in this unit cell. We have not explored this sort of variation.

## V. Discussion

The main aim of our theoretical computations was to definitely determine which experimental LEER spectrum corresponds to what number of MLs of hBN. We believe that the results of Section IV accomplish that goal. Those spectra that exhibit a single deep minimum, at 7 – 8 eV, are interpreted as arising from 1-ML hBN. Those spectra with a single, fully-formed minimum at about 2 eV (along with one or two minima at 8 – 12 eV) are interpreted as arising from 2-ML hBN. Thicker hBN can be identified by an increasing number of minima occurring in the 0 – 6 eV range, as in Fig. 1, although additional computations would in principle be necessary to fully understand those spectra i.e. in order to distinguish between a minimum due to an interlayer state between two hBN layers compared to a partial minimum arising from the downturn (as a function of decreasing energy) that is observed just above 0 eV.

Regarding the partial minimum just above 0 eV, we find analogous features at < 1 eV in many of the theoretical spectra already presented, for sufficiently small values of separation between the Ni surface and the adjacent BN layer. Fully formed minima at these very low energies are seen in Fig. 8 at 0.6 eV for the spectrum at 1.83-Å separation and 0.1-Å buckling, and in Fig. 9 at 0.8 eV for the spectrum at 2.08-Å separation and zero buckling. Partially formed minima are revealed by the downturn (as a function of decreasing energy) observed in many other theoretical and experimental spectra of Figs. 8 and 9. Such features at < 1 eV cannot by understood in terms of the type of interlayer states described in our prior work,[18,37] since the BN-Ni separation is much too small to support such states. In order to learn more about the origin of the features, we have studied how they evolve as a function of BN-Ni separation. The very low energy reflectivity features drop in energy as the BN-Ni separation increases, forming localized states with energies *below* the vacuum level. We monitor the energies of these states, and their associated pseudo-wavefunctions, for BN-Ni separations ranging from 1.8 to 4.0 Å (the actual physical separation for hBN on Ni occurs at the low end of this range, but we find it useful to explore the entire range in order to identify the nature of the states involved).

In Fig. 10 we plot, as a function of BN-Ni separation, the energies of reflectivity minima (for energies greater than the vacuum level) and of localized states related to those minima (for energies less than the vacuum level). Wavefunctions for a few selected states are shown, with results for many more states provided in the Supplemental Material. The wavefunctions are plotted as a function of the *z* distance perpendicular to the BN planes, and are averaged over the *xy* plane. To understand the eigenstates of the hBN-on-Ni system, we know from prior work that a suitable basis set is the one shown in Fig. 10(a): an hBN image-potential (IP) state on the left-hand side of the hBN, an interlayer (IL) state between the hBN planes (hBN IL), and another IL state between the hBN and Ni (BN-Ni IL). These IL states are themselves composed of combinations of image-potential states from the adjoining layers, as discussed at length in Ref. [37]. To this basis, we add a Shockley-type surface state of Ni, denoted Ni Sh. This type of state is well known for Ni and other transition-metal surfaces; for Ni(111) the Shockley state is located ~4 eV below the vacuum level (i.e. near the Fermi energy).[42,43,44,45]

There are three main branches in the energy plot of Fig. 10(b); let us start by considering the upper, full branch whose end points have wavefunctions shown in Figs. 10(c) and 10(d). These states are the "standard" type of interlayer states described in detail in our prior work, derived from image-potential states.[18,37] For Fig. 10(c) a peak in the wavefunction is seen between the two hBN layers, corresponding to an interlayer state in that space. The BN-Ni separation is much too small in this case to permit the formation of a BN-Ni IL state at this energy,[18,37] and hence a wavefunction peak is *not* seen between the hBN and the Ni. However, examining Fig. 10(d), we now see peaks both between the hBN planes and between the hBN and the Ni. Hence, the BN-Ni IL state has indeed formed for this relatively large BN-Ni separation, and the eigenstate pictured in Fig. 10(d) is composed of an admixture of BN-Ni IL state and the hBN IL state.

Turning to the lowest two branches of Fig. 10(b), we now have a situation that goes beyond what was discussed in our prior work. Considering large BN-Ni separations, two states are seen. The lowest energy one has energy and wavefunction (Fig. 10(h)) that agree well with a Shockley-type state of the Ni surface (see Supplementary Material, Fig. S5). As the BN-Ni separation decreases, this state rises in energy until it has an avoided crossing with the hBN IP state. For the smallest BN-Ni separations considered, the eigenstate in Fig. 10(e) is seen to have significant Ni Sh character (peak near the Ni surface) together with some admixture of the hBN IL state (peak between the hBN planes). It is this combined state that gives rise to the reflectivity minimum found in the theory just above 0 eV, and observed in the experiment by the downturn in the reflectivity as the energy approaches the mirror-mode onset. This very low energy state is thus seen to be derived from the Ni Shockley-type surface state.

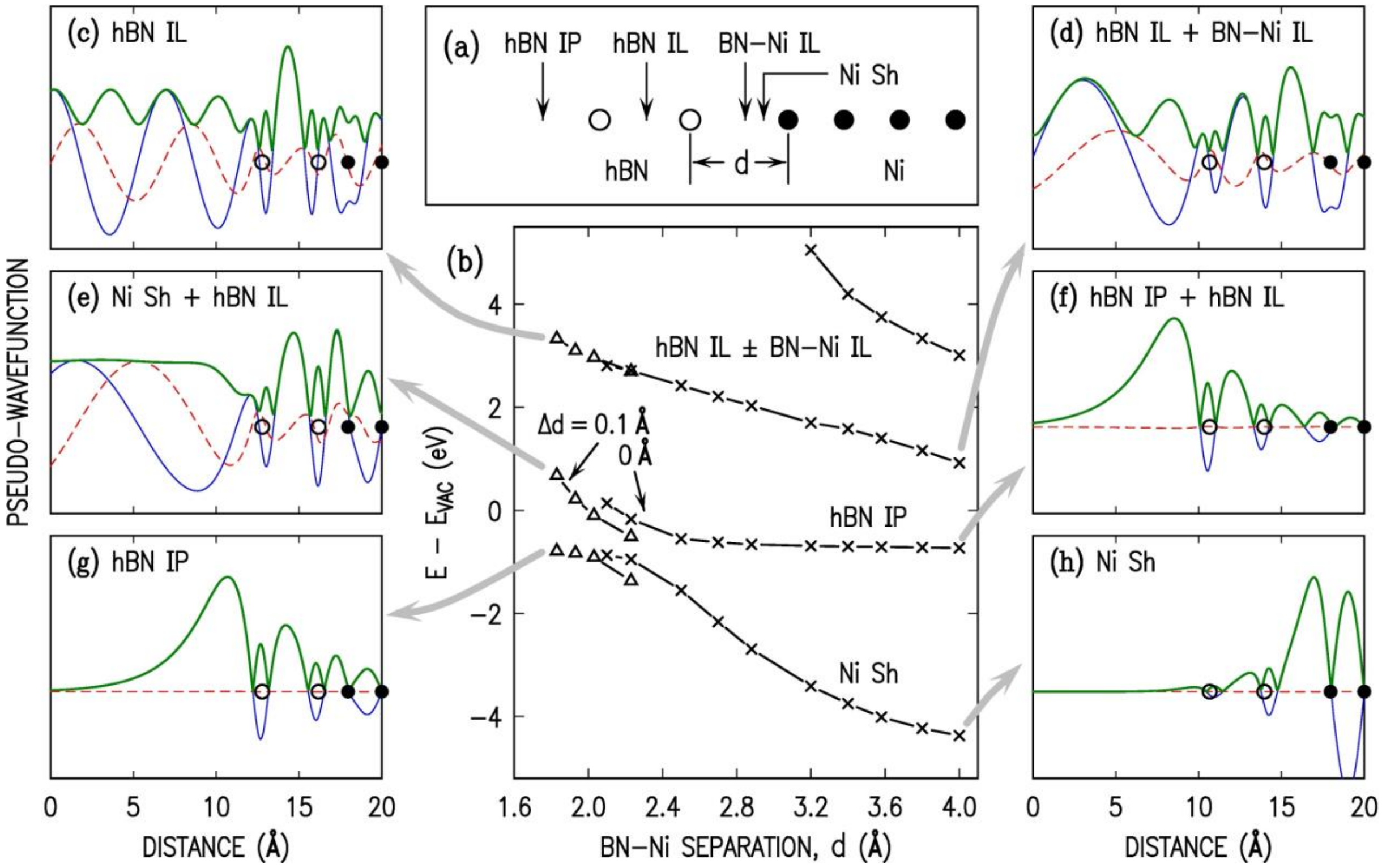

FIG 10. (a) Locations of basis states from which the complete eigenstates states of the hBN-on-Ni system are formed: hBN IP = image-potential state on left side of 2-layer hBN slab, hBN IL = interlayer state between hBN planes, BN-Ni IL = interlayer state between hBN and Ni surface, Ni Sh = Shockley-type state at surface of Ni slab. Closed circles denote Ni atoms, and open circles B atoms. (b) Energies of states relative to vacuum level $E_{VAC}$, for 2 layers of hBN on Ni. The energies of reflectivity minima are plotted for $E > E_{VAC}$, and of localized interlayer states for $E < E_{VAC}$. Energies for zero BN buckling are shown by x-marks, and for 0.1 Å buckling by triangles. The dominant character of the states is indicated. (c) – (h) Wavefunctions of specific states from (b), with character of each state (including admixtures) indicated. Real parts of the wavefunctions are indicated by thin solid (blue) lines, imaginary parts by dashed (red) lines, and magnitudes by thick solid (green) lines.

All of our computational results presented above have been for clean Ni surfaces, i.e. without the presence of oxygen or any other overlayer other than hBN. However, since our samples were transferred through air between the growth system and the LEEM, some oxidation of the Ni surface could occur. To investigate the possible influence of oxidation on our results, we have also considered the effect of an oxidized Ni surface (i.e. a NiO structure) between the hBN and the Ni substrate. The presence of such an oxide layer is very similar to that previously discussed for an oxidized Co surface, covered with hBN. The oxide produces a dipole at the interface, shifting the onset of the Ni NFE bands (i.e. Figs. 5(b) and 7(b)) from an energy of 2 – 3 eV above the vacuum level to an energy near or below the vacuum level.

An example is displayed in Fig. 11 for 2 ML of hBN on Ni(111) with a NiO layer terminating the Ni crystal. We employ a structure for the 1×1 NiO layer following Ref. [21], with the Ni plane being at a position as for bulk Ni, a monolayer of O atoms bonded at hollow sites of that Ni plane, and with the plane of O atoms vertically above the Ni plane by 1.3 Å. The average separation between the O plane and the overlying BN is taken to be 2.03 Å, with bulking of the BN of 0.1 Å (B atoms closer to the Ni plane than are the N). Comparing these results to those of Fig. 7, with no NiO layer, it is apparent that the presence of the plane of O atoms has relatively little effect on the final spectrum (i.e. once inelastic effects are included), despite the large shift of the Ni NFE bands seen in Fig. 11(b) compared to 7(b) produced by the dipole of the NiO layer. The final spectra of both Figs. 7 and 11 both show a clear minimum at 2 – 3 eV eV, arising primarily from the interlayer state between the two hBN layers. They also both show a broad minimum at 7 – 10 eV, arising from the higher hBN bands that couple to the interlayer band.

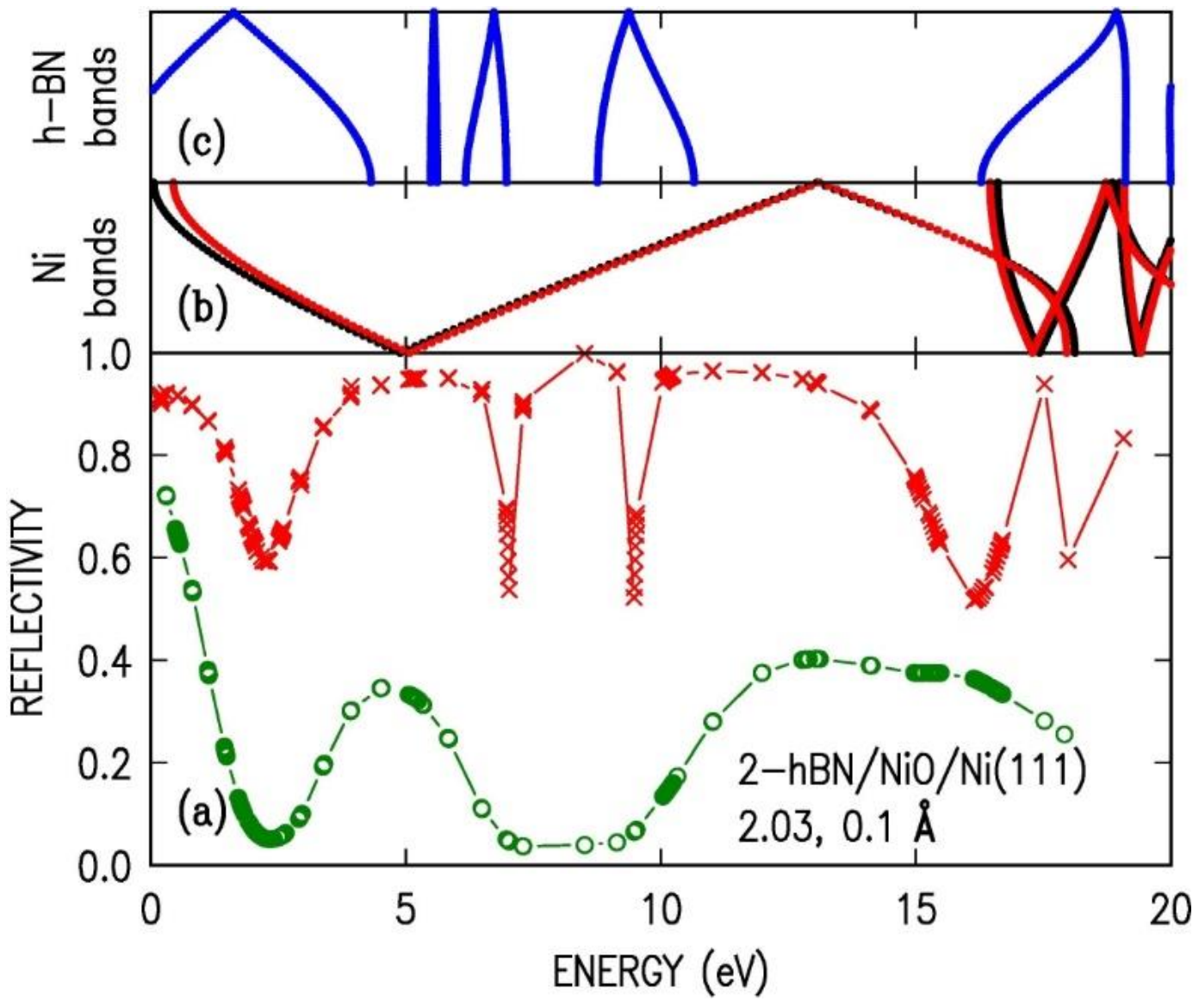

FIG 11. Same caption as Fig. 5, but for 2 ML of hBN on Ni(111) and a NiO layer between the hBN and the Ni. Average separation between the O plane of the NiO and the adjacent BN layer is 2.03 Å, and that BN plane is buckled by 0.1 Å (B atoms closer to the Ni plane than are the N).

In our prior publication, we incorrectly argued that an oxide layer *was* present on the hBN-covered Ni (or Co) surfaces, since there was no indication in the experimental spectra of a rise to unity reflectivity at an energy corresponding to the upper edge of the surface gap of the metal, i.e. as seen in Fig. 7(b) in the absence of inelastic effects.[24] However, we now realize that inclusion of inelastic effects is amply sufficient to inhibit any such rise to unity reflectivity, and hence the absence of such behavior certainly does not imply the presence of a NiO layer. Our samples may or may not be oxidized; our main point is that the reflectivity spectrum is rather insensitive to the presence of the O. However, this present result that the edge of the surface gap is not apparent in the reflectivity spectra should not be taken to be true in general. For example, in Ref. [15], for experimental spectra from a bare Cu surface, the edge of the surface gap *is* clearly apparent. The difference between that case and that of Figs. 7 and 11 is that the presence of 2 ML of hBN is apparently sufficient to entirely determine the reflectivity, independent of the location of the surface gap of the underlying Ni.

## VI. Conclusions

In summary, we have demonstrated how reflectivity spectra obtained in a LEEM can be used to identify the number of hBN monolayers grown on Ni, and with this information we have characterized the evolution of hBN films grown on polycrystalline Ni substrates. We find that there are a number of influences that determine the final arrangement of features in the LEER spectra: (i) The interlayer states associated with the "interlayer band" of hBN, located at 0 – 5 above the vacuum level, which for $n$ layers of hBN will lead to $n-1$ low-energy reflectivity minima. (ii) Replication of those low-energy minima by coupling between the interlayer band and the three bands immediately above it, located at 5 – 12 eV above the vacuum level. In practice this coupling is relatively weak (and it varies with buckling of the hBN), so that once inelastic effects are considered the additional features formed by these higher bands consist of just a single broad minimum at 7 – 10 eV. This broad minimum is an important feature in the LEER spectra, since it provides a means of chemical identification to distinguish between graphene and hBN (e.g. as might be encountered in a study which combines the two materials). (iii) A slight flattening of the experimental reflectivity at very low energies < 1 eV, which is due to the presence of a state located just below the vacuum level that is derived from the Ni(111) Shockley surface state.

**Acknowledgements**

This work was supported in part by the Center for Low Energy Systems Technology (LEAST), one of six centers of STARnet, a Semiconductor Research Corporation program sponsored by MARCO and DARPA. Discussions with Gong Gu are gratefully acknowledged, and we are grateful to Dacen Waters for critically reading the manuscript.

[41] W. Paszkowicz, J.B. Pelka, M. Knapp, T. Szyszko, S. Podsiadlo, Appl. Phys. A **75**, 431 (2002).
[42] A. Goldmann, V. Dose, and G. Borstel, Phys. Rev. B **32**, 1971 (1985).
[43] G. Borstel, G. Thörner, M. Donath, V. Dose, and A. Goldmann, Solid State Comm. **55**, 469 (1985).
[44] F. Mittendorfer, A. Eichler, J. Hafner, Surf. Sci. **423**, 1 (1999).
[45] L. Maguad, A. Pasturel, P. Mallet, S. Pons, and J.-Y. Veuillen, Europhys. Lett. **67**, 90 (2004).

**Characterization of hexagonal boron nitride layers on nickel surfaces by low-energy electron microscopy**

P. C. Mende, Q. Gao, A. Ismach, H. Chou, M. Widom, R. Ruoff, L. Colombo, and R. M. Feenstra

## <u>Supplementary Material</u>

Pseudo-wavefunctions, as computed by VASP using the projector augmented wave technique,[1] are shown below for various arrangements of hBN and Ni. All wavefunctions are plotted as a function of the $z$ distance perpendicular to the BN planes, and are averaged over the $xy$ plane. States are labeled according to their energies $E$ relative to the vacuum level, $E_{VAC}$, for each situation. For in-plane wavevector of zero, states with $E > E_{VAC}$ are propagating states in the vacuum, formed in accordance to our reflectivity theory with an incident plane wave and a reflected wave on the left side of the hBN, and a transmitted wave on the right traveling into the vacuum or into the Ni. States with $E < E_{VAC}$ are non-propagating states ($k_z = 0$), decaying into the vacuum region on both sides of the slab. These states occur in pairs, with symmetric (even) or antisymmetric (odd) parity relative to the midpoint of the slabs. (The states with $E > E_{VAC}$ do not have definite parity, since they are propagating waves). All wavefunctions are plotted using thin solid (blue) lines for to the real part, thin dashed (red) lines for the imaginary part, and heavy solid (green) lines for the magnitude. Positions of B atoms on the plots are shown by open circles, and of Ni atoms by filled circles.

## I. Wavefunctions for 2 Layers of hBN with typical separation to Ni

Figure S1 shows wavefunctions corresponding to Fig. 7 of the main text. There are two distinct minima that occur in the LEER spectrum of Fig. 7 including inelastic effects: one near 3 eV and the other at 7 – 10 eV relative to the vacuum level. The latter minimum is seen in the computations that neglect inelastic effects to consist of two minima, near 7 and 9 eV. We consider states near the lowest reflectivity values for all these minima, namely, at 2.87, 7.23, and 9.35 eV relative to the vacuum level. Inspecting the wavefunctions of the various states, we see that the one at 2.87 eV has most of its weight located in the interlayer space between the two hBN layers. In contrast, the states at 7.23 and 9.35 eV have more weight on or near the atomic planes of hBN. The atomic orbitals from which these latter two states are derived have substantial $p_z$ character,[2] and hence the wavefunctions (for the 7.23-eV state in particular) have little weight *exactly* on the BN atomic planes; nevertheless, they have substantially more weight near those planes than does the 2.87-eV state.

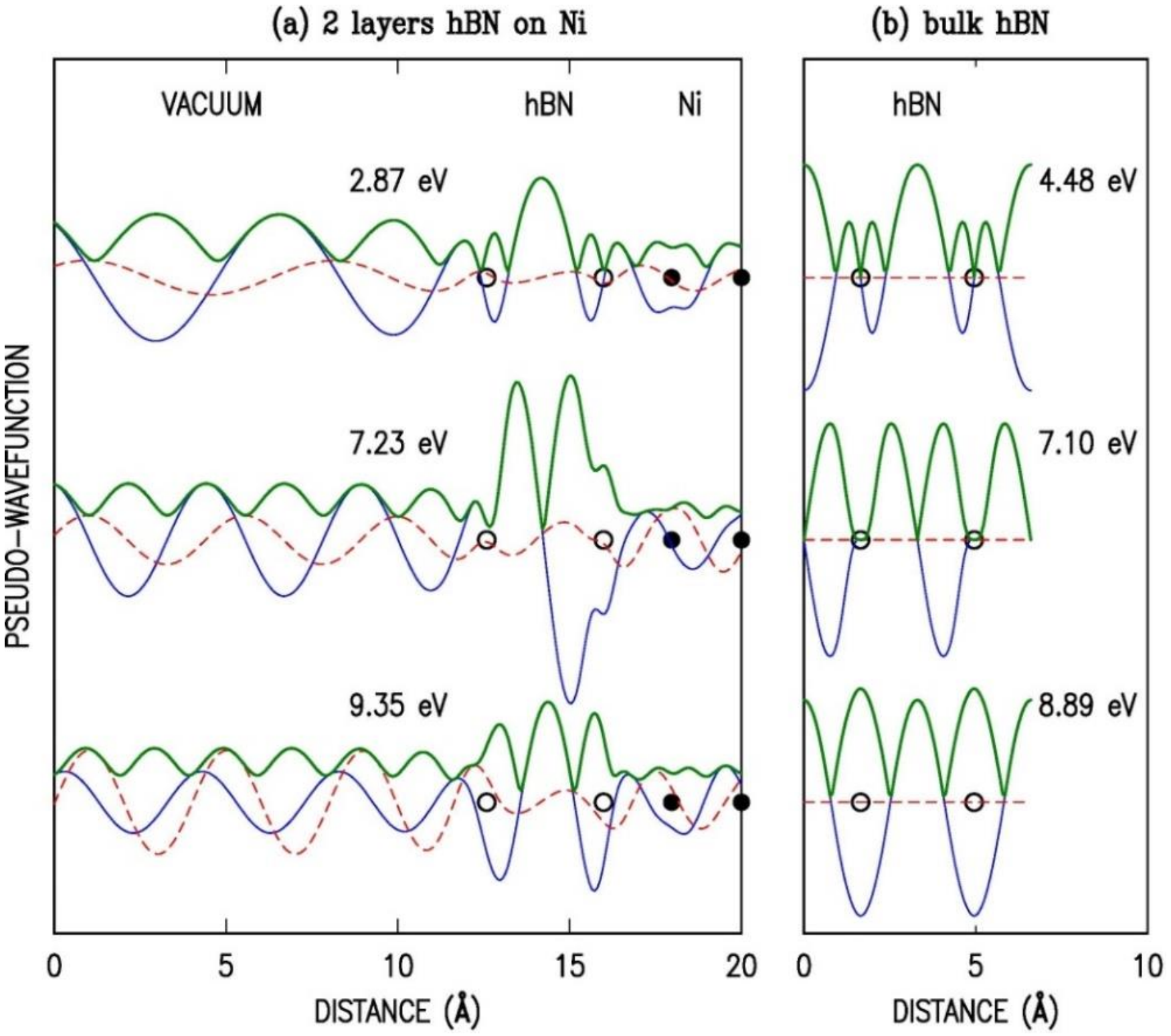

FIG S1. Pseudo-wavefunctions for (a) 2 layers hBN on Ni with 2.03 Å separation between Ni and the adjacent hBN layer and 0.1 Å buckling of that layer (B atoms closer than N to the Ni), and (b) bulk hBN. Wave-functions in (a) are normalized to an incident plane wave from the left, with unit amplitude.

For comparison with the hBN-on-Ni wavefunctions of Fig. S1(a), we plot in Fig. S1(b) corresponding states of bulk hBN. We consider the specific bulk bands that give rise to the respective reflectivity minima, and we choose states within the bulk band at the band extrema, with $k_z$ values of zero (such states have no $\exp(ik_z z)$ variation, thus simplifying the comparison between the results of Figs. S1(a) and (b)). We see quite good agreement in the overall shape of the states in Fig. S1(b) with those of (a), in particular for the large weight in the interlayer space of the lowest energy state and the nodal structure of the higher energy states.

**II. States of free-standing 2-layer hBN slab**

Figure 10 of the main text shows energies and wavefunctions for various interlayer states of hBN on Ni. In order to understand the character of those states, we show in this and the following Section states for free-standing slabs of hBN and of Ni. Figure S2 shows wavefunctions for a free-standing 2-layer hBN slab. The state at 2.06 eV occurs at a minimum of the reflectivity of the slab. The states at –0.49 and –0.95 eV correspond, respectively, to antisymmetric and symmetric combinations of the two image-potential states that exist on either side of the slab (all states associated with the BN conduction band occur at $E > E_{VAC}$ for this 2-layer slab). To obtain the IP(L) and IP(R) states separately, we can take the sum and difference of the antisymmetric and symmetric states shown in Fig. S2; the energy of the separate states would be about $-0.72$ eV.

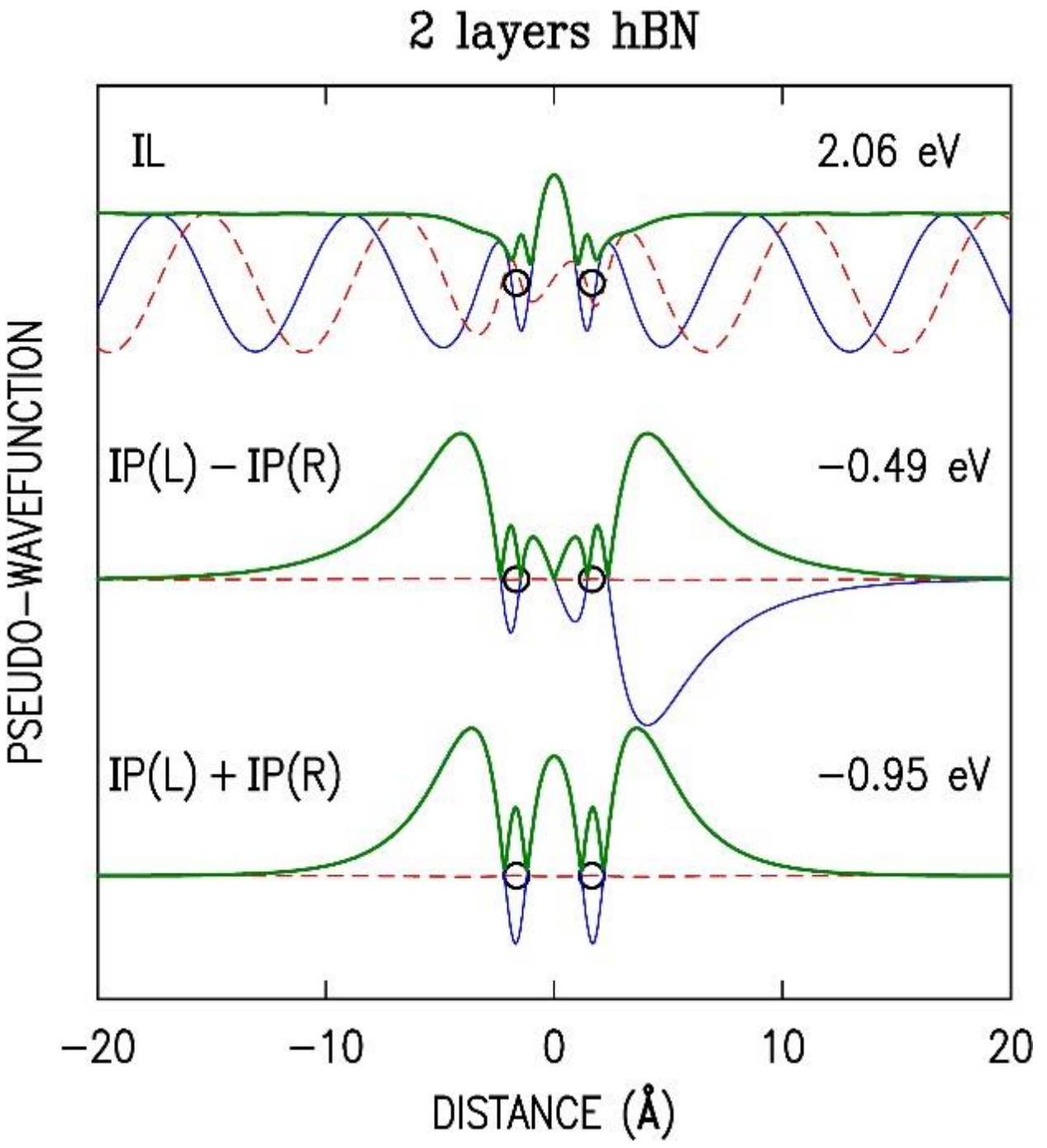

FIG S2. Pseudo-wavefunctions for 2 layers of free-standing hBN. The upper curve shows the interlayer (IL) state between the two hBN planes, whereas the lower two curves show antisymmetric (–) and symmetric (+) combinations of image-potential (IP) states on either side of the hBN slab. The IP state on the left side of the slab is by denoted IP(L), and the one on the right by IP(R).

## III. States of free-standing Ni slabs

Figure S3 shows energies and wavefunctions for majority-spin states of free-standing slabs of Ni, considering slabs with thickness of 3, 5, and 7 layers. In the band structures of Figs. S3(a) and (b), the image-potential (IP) and the Shockley-type (Sh) surface states are easily identified. At the Γ-point, the IP states have energies just below $E_{\mathrm{VAC}}$ (they are the only states with energies slightly below $E_{VAC}$). For $E > E_{VAC}$, there are two kinds of states, those with large amplitude in the vacuum and small amplitude in the Ni, and those with substantial amplitude in both regions. The former arise from free-electron type states in the vacuum (the amplitude of these states in the Ni goes to zero for large vacuum widths), whereas the latter are associated with the nearly-free electron (NFE) band of bulk Ni. The onset of the Ni NFE band can be determined by matching the *xy*-averaged potentials of the slab and a bulk Ni computation (i.e. as done in the main text for aligning the bulk bands in panel (b) of Figs. 5, 7, and 11). For the 3-layer Ni slab of Fig. S3(a), the onset of the NFE band of Ni is found to occur at 0.69 eV relative to $E_{VAC}$, and for the 7-layer slab it is at 0.60 eV. For the Sh states, they exist in the gap that occurs between the $d$-band (energies of about $-6$ to $-8$ eV relative to $E_{VAC}$) and the NFE band. The specific states of the 7-layer slab of Fig. S3(b) identified as the Sh states are in good agreement with the prior work of Magaud et al. (the symmetric-antisymmetric splitting is 0.34 eV for our 7-layer slab, whereas it is 0.2 eV for the 9-layer slab of those workers).[3] For the 3-layer results of Fig. 10(a), the lower (symmetric) Sh state moves down into the $d$-band states, for wavevector at the Γ-point. Nevertheless, this symmetric Sh state can still be easily identified in the computations since its $xy$-averaged wavefunction has a nonzero result, in contrast to that obtained for all nearby $d$-band states which produce a zero $xy$-averaged wavefunction due to their $d_{xz,yz}$ character.

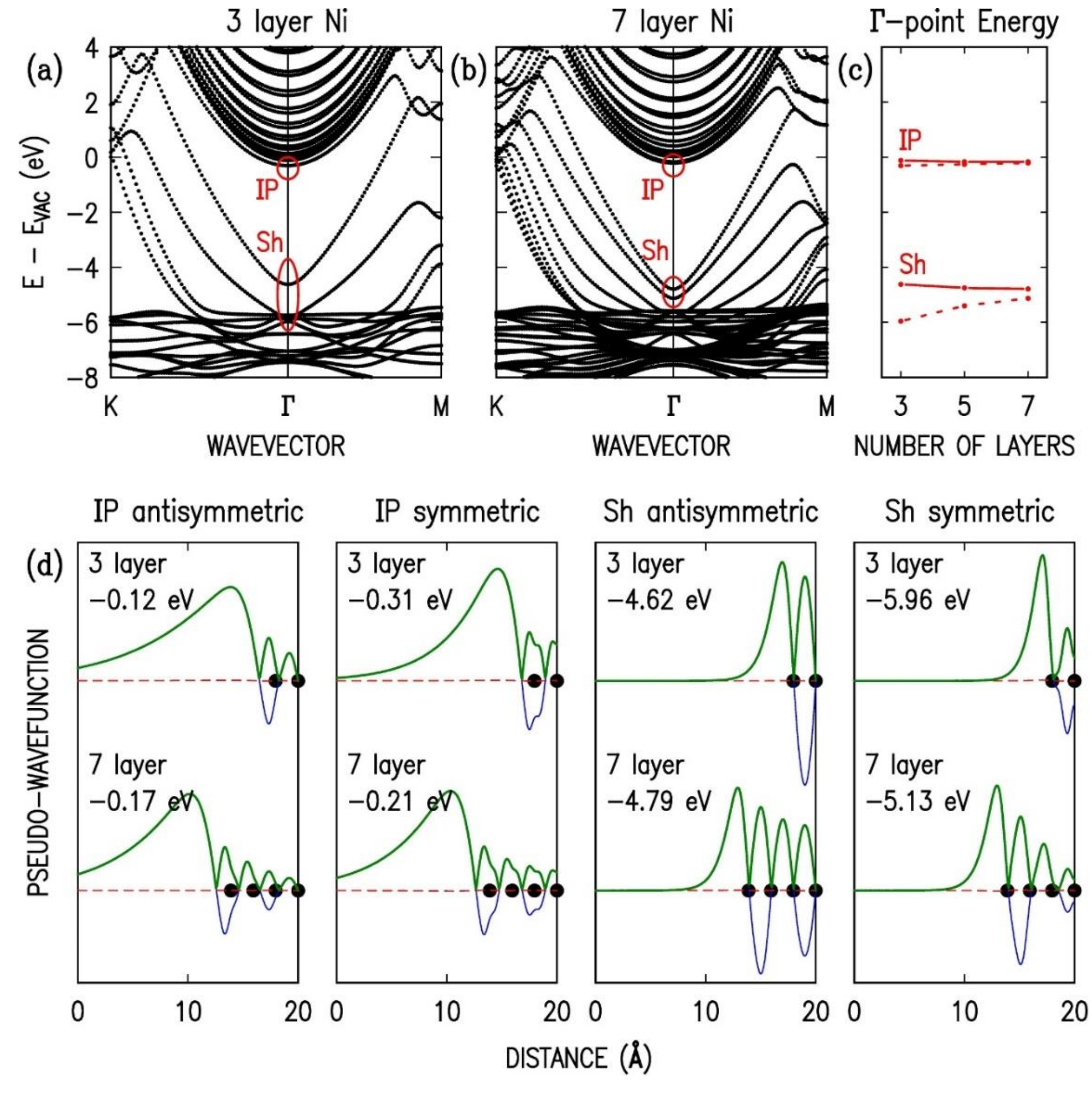

FIG S3. (a) and (b) Band structures of 3- and 7-layer Ni slabs, with image-potential (IP) and Shockley-type (Sh) states indicated. (c) Summary of Γ-point energies of the IP and Sh states; solid lines are used for antisymmetric states, and dotted lines for symmetric states. (d) Pseudo-wavefunctions for the IP and Sh states.

The very different character of the IP and Sh states is evident in Fig. S3(d): For the lobes of the wavefunctions extending into the vacuum, the Sh states are localized much closer to the Ni surface than are the IP states. Also, the Sh states are seen to have nodes at the locations of Ni atoms, whereas the IP states have a shorter wavelength in the Ni and, for the case of the 7-layer slab, nearly display a maximum at the location of the surface Ni atom.

### IV. Wavefunctions for hBN on Ni

In Fig. 10 of the main text, wavefunctions are displayed at a select few BN-Ni separations for the hBN-on-Ni system. In Fig. S4, we show wavefunctions for nearly all of the separation values considered in Fig. 10(b) (both Figs. 10 and S4 display antisymmetric states for $E < E_{VAC}$). Figures S4(a) –(c), respectively, shows states corresponding to the uppermost full branch of Fig. 10(b), to

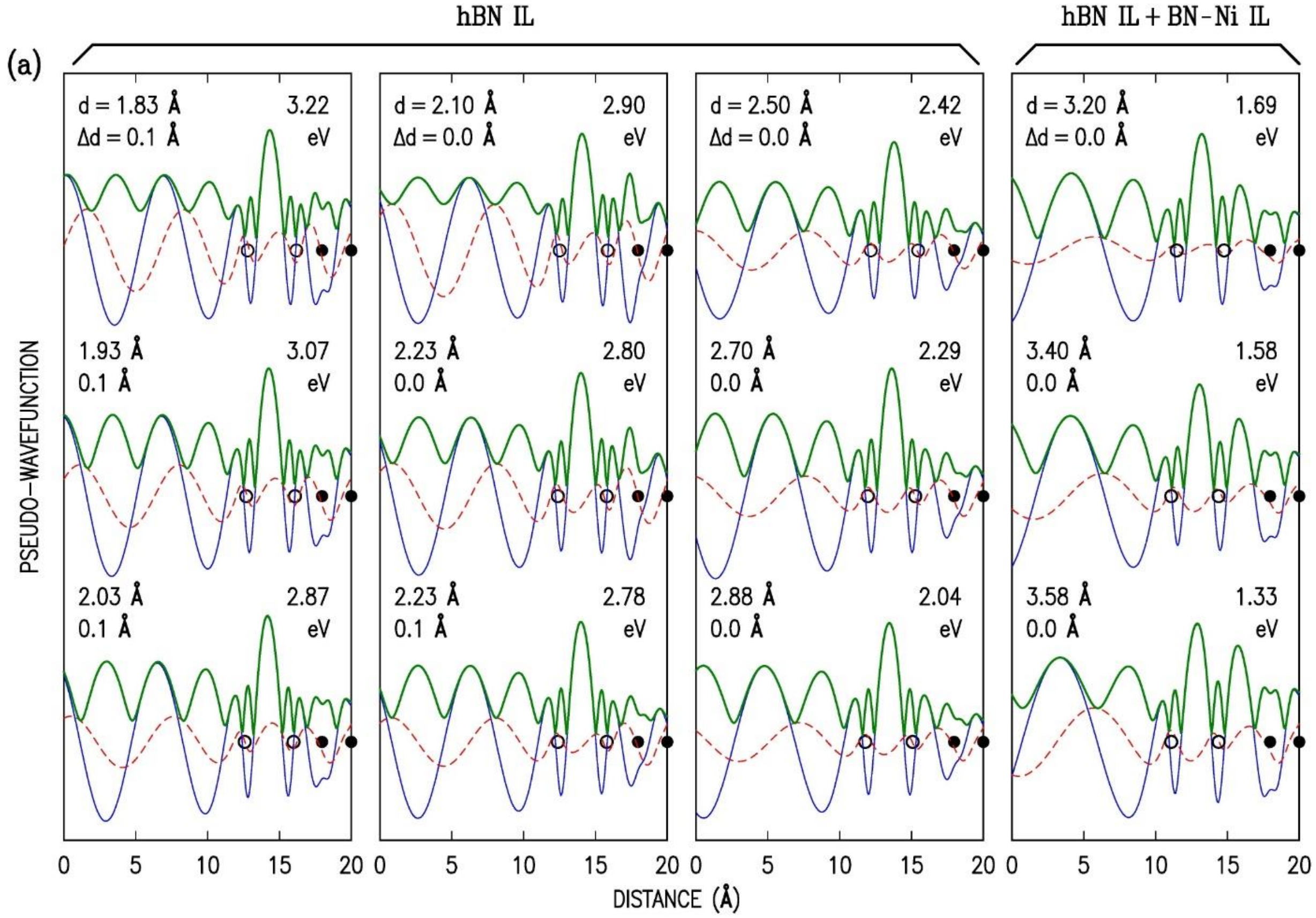

(b)

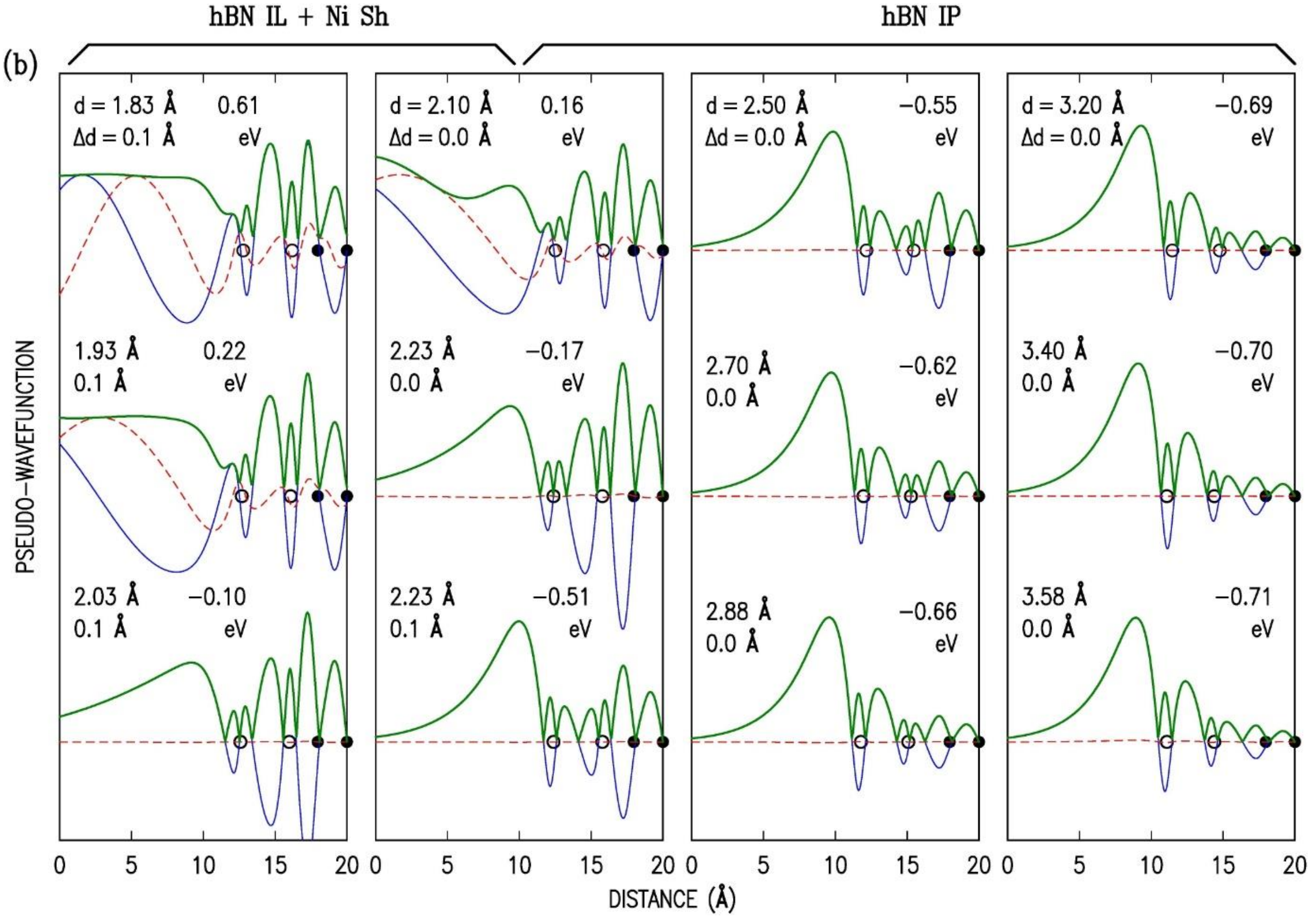
hBN IL + Ni Sh
hBN IP
d = 1.83 Å 0.61
Δd = 0.1 Å eV
1.93 Å 0.22
0.1 Å eV
2.03 Å −0.10
0.1 Å eV
d = 2.10 Å 0.16
Δd = 0.0 Å eV
2.23 Å −0.17
0.0 Å eV
2.23 Å −0.51
0.1 Å eV
d = 2.50 Å −0.55
Δd = 0.0 Å eV
2.70 Å −0.62
0.0 Å eV
2.88 Å −0.66
0.0 Å eV
d = 3.20 Å −0.69
Δd = 0.0 Å eV
3.40 Å −0.70
0.0 Å eV
3.58 Å −0.71
0.0 Å eV
PSEUDO-WAVEFUNCTION
DISTANCE (Å)

(c)

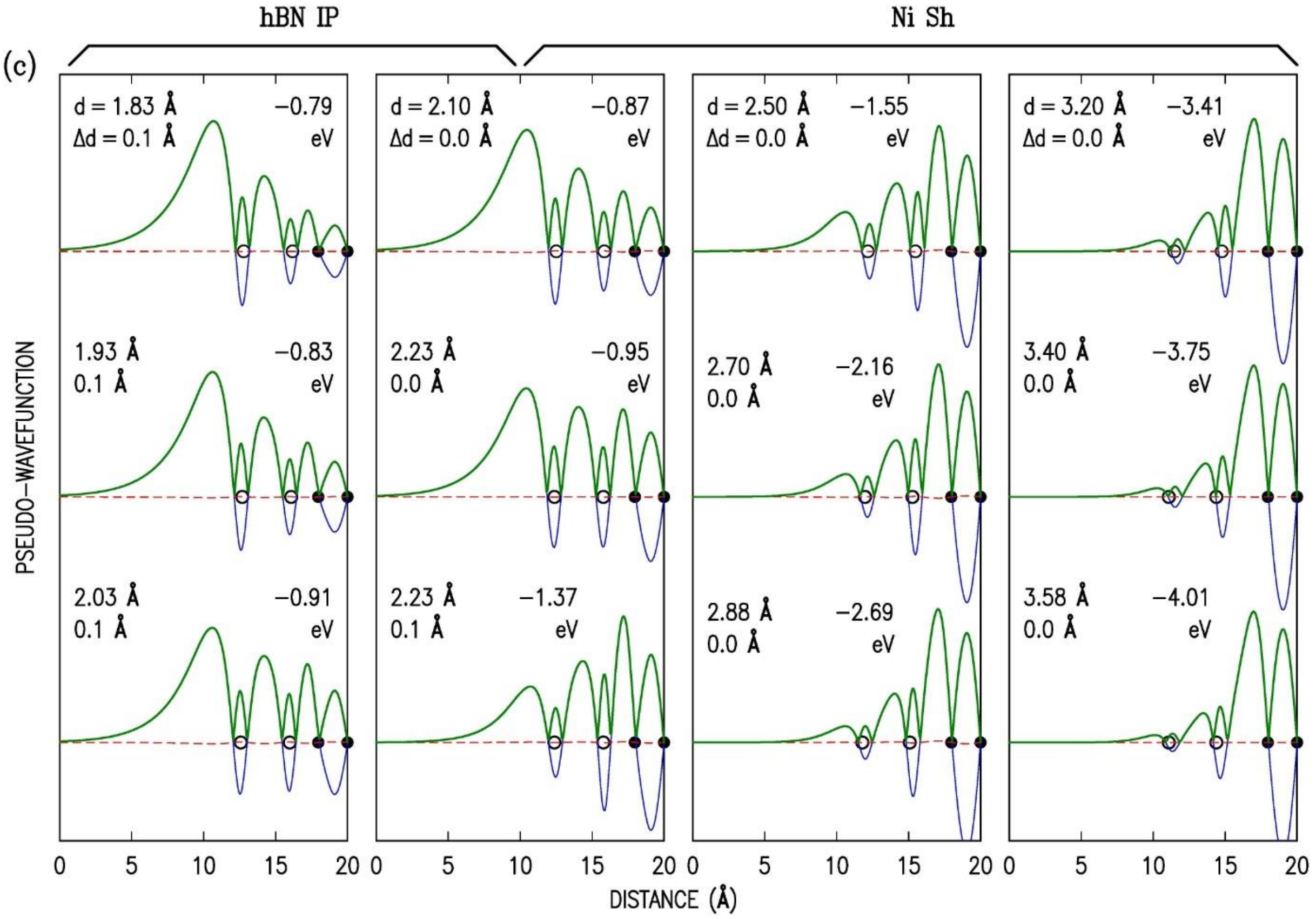
hBN IP
Ni Sh
d = 1.83 Å −0.79
Δd = 0.1 Å eV
1.93 Å −0.83
0.1 Å eV
2.03 Å −0.91
0.1 Å eV
d = 2.10 Å −0.87
Δd = 0.0 Å eV
2.23 Å −0.95
0.0 Å eV
2.23 Å −1.37
0.1 Å eV
d = 2.50 Å −1.55
Δd = 0.0 Å eV
2.70 Å −2.16
0.0 Å eV
2.88 Å −2.69
0.0 Å eV
d = 3.20 Å −3.41
Δd = 0.0 Å eV
3.40 Å −3.75
0.0 Å eV
3.58 Å −4.01
0.0 Å eV
PSEUDO-WAVEFUNCTION
DISTANCE (Å)

FIG S4. Pseudo-wavefunctions for 2 layers of hBN on Ni, with average separation between Ni and the adjacent BN layer indicated by $d$ and the buckling of that layer indicated by $\Delta d$ (positive $\Delta d$ corresponds to B atoms closer than N to the Ni). (a) – (c) shows states from the various branches of the energy vs. separation plot of Fig. 10(b).

the next lowest branch, and to the bottom branch. The character of each state is identified in accordance to the basis set shown in Fig. 10(a) of the main text. Specifically, in Fig. S4(a), there is a peak in the wavefunctions between the two hBN planes, corresponding the hBN IL state. Additionally, for the 2 or 3 states with largest BN-Ni separations, a peak also is seen between the hBN and the Ni, corresponding to the BN-Ni IL state. The net state is thus an admixture, hBN IL + BN-Ni IL, with the BN-Ni IL state coming down in energy (uppermost branch of Fig. 10(b)) to approach the hBN IL branch for large BN-Ni separations.

For Fig. S4(b), the 7 states with the largest BN-Ni separation are seen to have primarily hBN IP character, i.e. with wavefunctions that agree well with an image-potential state on the left-hand side of the hBN layers (compare to Fig. S2, e.g. taking the sum of the wavefunctions for the states at −0.49 and −0.95 eV in order to form the IP(L) state). The 5 or 6 states with smallest BN-Ni separation display a substantial peak between the Ni and the hBN. This peak clearly arises from the Ni Sh state, whose energy is seen in Fig. 10(b) to rise up and cross that of the hBN IP state. For the 2 states with smallest BN-Ni separation in Fig. S4(b), at 0.22 and 0.61 eV, they are high enough in energy to admix with the higher lying hBN IL state, resulting in additional wavefunction peaks between the hBN planes.

For Fig. S4(c), it is the 5 states with smallest BN-Ni separation that display the dominant wavefunction lobes on the left-hand side of the hBN layers, i.e. signifying hBN IP character. For the 7 or so states with higher BN-Ni separation, they all display the dominant peak between the hBN and Ni, indicative of Ni Sh character. These states for very weakly bonded hBN on Ni are further examined in the following Section, to definitely demonstrate the similarity in character between them and the corresponding states of bare Ni.

## IV. Shockley-type states for hBN on Ni compared to those for bare Ni

A new aspect of our work is the identification of the very low-energy features (just above the mirror-mode transition) seen in both experimental and theoretical LEER spectra of hBN on Ni as arising from interlayer states that are derived from Shockley-type (Sh) surface states of the bare Ni. This identification is made on the basis of the energies shown in Fig. 10(b), since the lowest-energy branch there approaches the energy of the Sh state of bare Ni in the limit of large BN-Ni separation. Additionally, in Fig. S5, we explicitly compare the wavefunctions of the Sh states of bare Ni with the corresponding states for very weakly bonded hBN on Ni (separations of 3.58 and 4.0 Å). It can be seen that the wavefunctions for the very weakly bonded hBN on Ni and the bare Ni are very similar. Some modest confinement of the state occurs between the hBN and Ni, thus increasing its energy slightly, but other than that all details of the wavefunctions are in good agreement. This agreement definitively confirms our identification that these states for the hBN-on-Ni system originate from the Ni Sh state. Considering the Ni Sh state and the hBN IP(R) state to be a basis for forming these eigenstates of hBN-on-Ni, very little mixing occurs between the two basis states since the energy of the IP(R) state is so much higher (~ 4.6 eV neglecting confinement in the BN-Ni space, and > 6 eV with confinement) than that of the Ni Sh state.

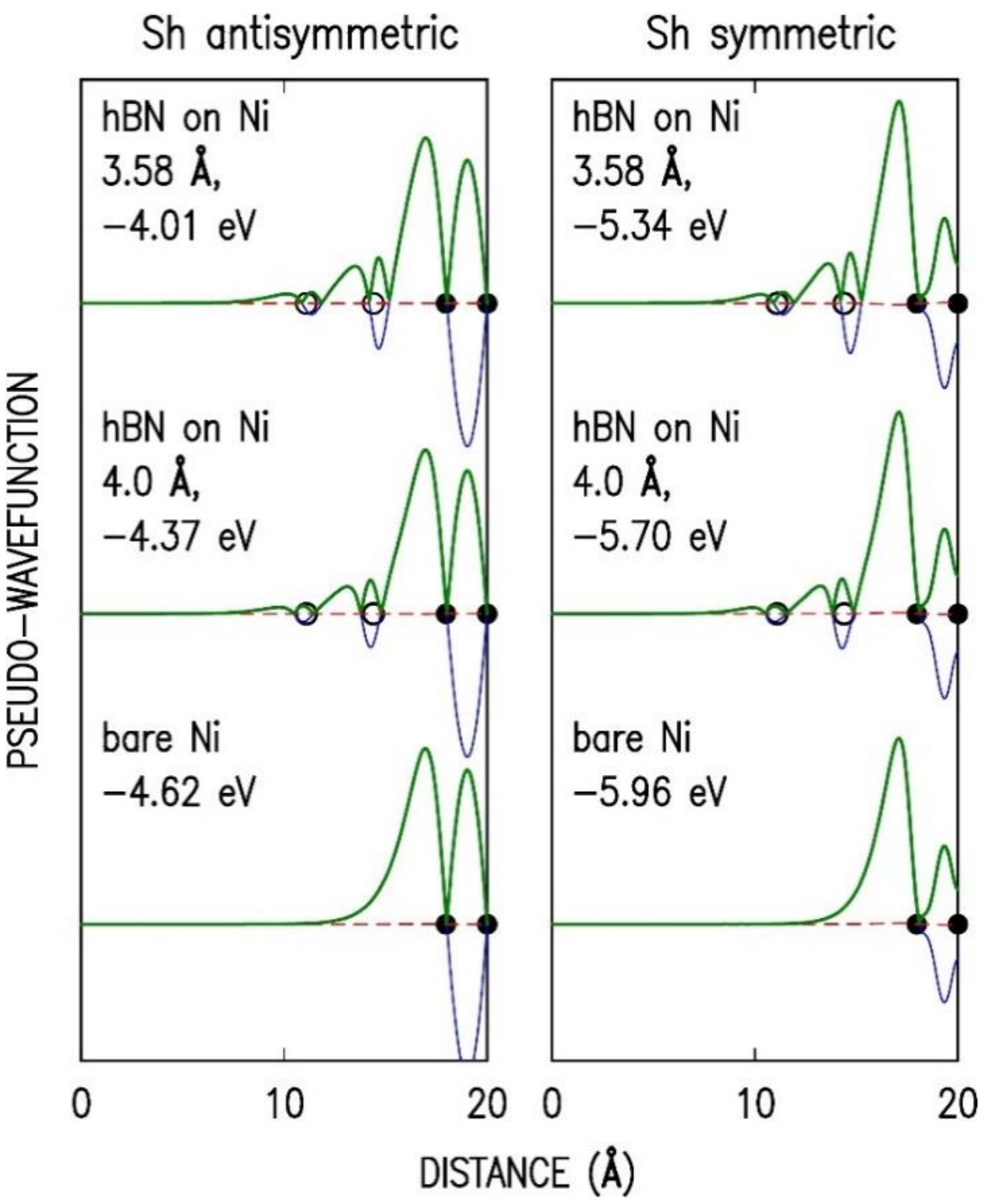

FIG S5. Pseudo-wavefunctions for Shockley-type states, comparing the case of 2 layers of hBN on Ni with Ni-BN separations 3.58 Å (top) and 4.0 Å (middle), with that of bare Ni (bottom). Thickness of the Ni slab is 3 layers in all cases, and states that are antisymmetric and symmetric relative to the center of the Ni slab are displayed. For the particular states chosen for hBN on Ni, they are the ones whose energies (and wavefunctions) approach that of the Shockley state of bare Ni, in the limit of large Ni-BN separation.

## V. LEER spectra for 2 Layers of hBN weakly bonded to Ni

To complete our discussion of the influence of BN-Ni separation on the LEER spectra, we show in Fig. S6 the computed spectra for 2 ML of hBN weakly bonded to Ni, with average BN-Ni separation of 3.58 Å and buckling of zero. We are focusing now on energies above $E_{VAC}$, i.e. as probed by LEER (unlike the prior Section in which we studied energies far below $E_{VAC}$ in order to understand the nature of particular states there). For this large BN-Ni separation, we expect *two* reflectivity minima associated with interlayer states, since there is one such state associated with the interlayer space between the Ni and the first hBN layer, and another in the space between the two hBN layers (both of these interlayer states are composed of combinations of image-potential states from the adjacent layers). Two linear combinations of these two IL states are formed, with the bonding combination ending up with lower energy than the antibonding one. This type of expectation follows our prior results for graphene on Cu (which is a weakly bonded system, with Cu-graphene separation of about 3.3 Å),[4] and indeed, the results of Fig. S6 are consistent with the presence of two IL states. However, a slightly detailed explanation of the computed spectra is required in order to delineate the two states.

First consider the spectrum of Fig. S6 that neglects inelastic effects (shown by red x-marks). This spectrum shows a clear minimum at 3.75 eV. Additionally, a downturn as a function of decreasing energy is evident, at about 1.6 eV. However, for lower energies the reflectivity then abruptly rises to unity, due to the presence of the gap in the spectrum of Ni states. If this gap did not occur, then we would expect a second reflectivity minimum at an energy slightly less than 1.6 eV, but this second minimum is not visible due to the presence of the gap. Nevertheless, if we turn to the spectrum that includes inelastic effects (green circles), we now find a distinct minimum at 1.40 eV. We associate this minimum with the lowest energy one, i.e. the "missing" one from the elastic-only computation. (Inelastic effects are seen to serve here to largely obscure the presence of the

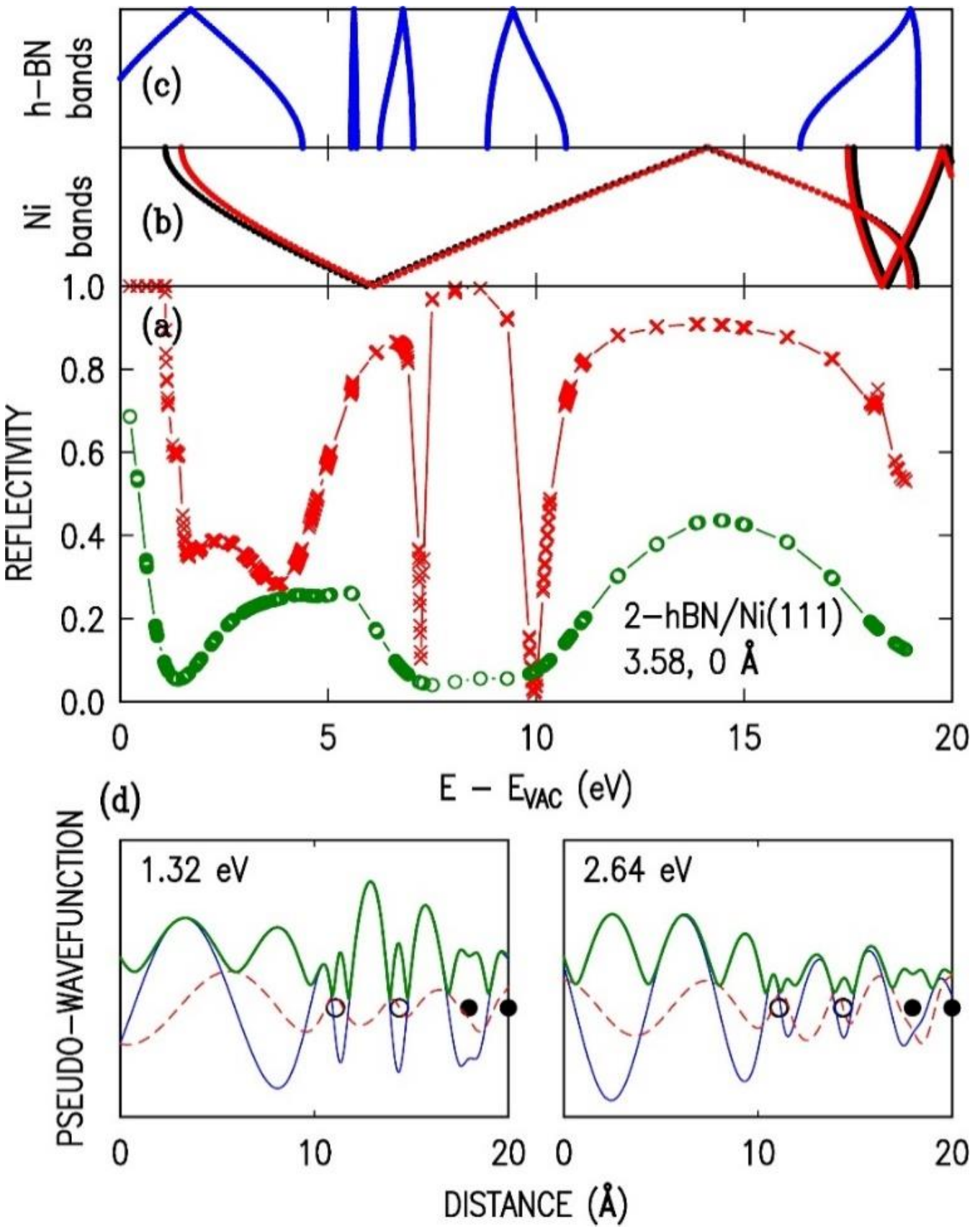

FIG S6. (a) Computed reflectivity spectra for 2 ML hBN on Ni(111), with (circles) and without (x-marks) inelastic effects, averaged over minority and majority spins. Average Ni-BN separation is 3.58 Å and buckling is zero. (b) Bulk Ni band structures in (111) direction (majority spin band has noticeably lower energies than the minority spin band, for energies < 5 eV). (c) Bulk hBN band structure in (0001) direction. (d) Pseudo-wavefunctions for states with energies near the reflectivity minima.

Ni band gap, consistent with the discussion in the main text surrounding the possible role of oxygen at the hBN/Ni interface). However, in this spectrum that includes elastic effects, the higher energy minimum at 3.75 eV is no longer visible. Apparently, inelastic effects are sufficient to obscure that minimum. In any case, examining the wavefunctions for states near these two minima, Fig. S6(d), we do indeed find the expected behavior. For the 1.32-eV state, clear peaks in the wavefunctions are seen both between the hBN planes and between the hBN and the Ni, indicative of IL states at both locations. For the 2.64-eV state, peaks are again seen at these locations, but these peaks are relatively weak; apparently this state is only a weak resonance (and hence, it is not surprising that it was obscured from the reflectivity spectrum when inelastic effects were included). In any case, we do indeed find two resonant IL states for this weakly bonded situation, precisely as expected based on our prior theory of IP states combining into the IL states.[4,5] For the present case of 3.58-Å separation, the upper resonance is rather weak. However, for separations of about 3.8 Å or larger the upper state is stronger, and it is directly revealed in spectra that include inelastic effects.